\documentclass[useAMS,usenatbib]{mn2e}
\usepackage{graphicx}
\usepackage{amsmath}
\DeclareMathOperator\erf{erf}

\def\Cplusplus{C\raisebox{0.5ex}{\tiny\textbf{++}}}

\title[Outer and inner mass distributions of UGC 4284 and UGC 11861]
{Outer and inner mass distributions of the irregular galaxies UGC 4284 and UGC 11861: Constraining the baryonic content through stellar population synthesis studies}

\author[P. Repetto, et al.]{P. Repetto$^{1}$\thanks{E-mail:prepetto@lna.br}, 
Eric E. Mart\'{\i}nez-Garc\'{\i}a$^{2}$, M. Rosado$^{3}$, R. Gabbasov$^{4}$\\
$^{1}$Laborat\'{o}rio Nacional de Astrof\'{i}sica, Rua Estados Unidos 154, 37504-364, Itajub\'{a}, MG, Brazil\\
$^{2}$Instituto Nacional de Astrof\'{\i}sica \'Optica y Electr\'onica, Luis Enrique Erro 1, Tonantzintla, Puebla, M\'{e}xico C.P. 72840\\
$^{3}$Instituto de Astronom\'{\i}a, Universidad Nacional Autonoma de M\'{e}xico, Circuito de la Investigaci\'on Cient\'{\i}fica, \\
Ciudad Universitaria, M\'{e}xico, D.F., C.P. 04510\\
$^{4}$Instituto de Ciencias B\'asicas e Ingenier\'{\i}as, U.A.E.H., Carretera Pachuca-Tulancingo, M\'{e}xico, C.P. 42184}
\begin{document}

\date{}

\pagerange{\pageref{firstpage}--\pageref{lastpage}} \pubyear{2002}

\maketitle

\label{firstpage}

\begin{abstract}

In this article we investigate the outer and inner mass distributions of the irregular galaxies UGC 4284 and UGC 11861, taking advantage of published HI and 
H$\alpha$ high resolution rotation curves and constraining the { stellar} disk of both galaxies throughout stellar population synthesis studies. { In addition
we take into account the gas content of both galaxies deriving the HI$+$He rotation curve.} The deduced baryonic rotation curves { (star+gas)} 
are inadequate to account for the total mass of UGC 4284 and UGC 11861, for that reason we examine the possibility of dark matter to explain the incongruity 
between the observed HI and H$\alpha$ rotation curves of UGC 4284 and UGC 11861 and the derived baryonic rotation curves.
{ We consider NFW, Burkert, DiCintio, Einasto, and the Stadel dark matter halos, to analyse the dark matter content of UGC 4284 and 
UGC 11861. The principal results of this work are that cored dark matter models better reproduce the dark matter H$\alpha$ and HI rotation curves of UGC 11861 and 
the dark matter HI rotation curve of UGC 4284, while, the H$\alpha$ rotation curve of UGC 4284 is better reproduced by a cuspy DiCintio DM model. In general, cored 
exponential two-parameters models Einasto and Stadel, give better fits than Burkert.} This trend, as well as to confirm past results, presents for the first time
a comparison between two different exponential dark matter models, Einasto and Stadel, in an attempt to better constrain the range of possible exponential dark matter 
models applied to real galaxies. 

\end{abstract}


\section{Introduction}~\label{intro}

The rotation of gas and/or stars in galaxies provides an important information about the kinematics of the
galactic disk. Observed RCs are somewhat flat and no Keplerian decline was detected. The interpretation based on the dark matter (DM) paradigm assumes a 
large amount of DM far beyond the visible part of galaxies. \citet{Freeman1970} was one of the first to suggest that RCs could be used as indication of DM, 
reporting that the RCs of NGC 300 and M33 did not show the expected Keplerian decline, and estimating that the DM mass in these galaxies should be at least 
as large as the mass of the luminous matter. \citet{Rubin1980} and \citet{Bosma1978, Bosma1981a, Bosma1981b} endorse the evidence of DM, performing extensive studies of 
galaxies RCs in optical and radio wavelengths. { \citet{Bahcall1982} built artificial RCs considering a central component, a disk, a spheroidal component
and a DM component to determine unexpected deviations from the generally believed logarithmic slope of 2.0. \citet{Carignan1985} analysed the mass distribution
of four galaxies detecting the necessity of a DM component to explain the observed RCs of those galaxies whose RCs extend to $1.5\,-\,3.0\, R_{25}$.} 
These works turn out a general acceptance of the DM hypothesis.\\ 

In general, the first step to determine the DM distribution in galaxies is to recognize what kind (bulge, disk, and bar) and how many stellar 
components constitute the innermost optical disk. This can be done by performing isophotal analysis on broad band image photometry of the studied objects. 
In fact, the luminosity profile obtained from galaxies observations, is the unique information to derive a coarse estimate of the visible mass of such objects, 
assuming a constant relation between luminosity (L) and mass (M). Once all the stellar components are identified, different mass profiles for the stellar 
disk, bulge, and bar are used to account for the stellar mass of the optical disk. The gas mass content is represented by the mass profile of the gaseous tracer 
(usually neutral hydrogen) used to measure the RC far from the optical disk extent. The DM halo is commonly considered spherically symmetric and its circular 
velocity is given by the ordinary Newtonian law. The next step is to accomplish a non-linear least squares fit to the observed RC considering a general fitting 
relation $V^2_{tot}(R)=\left[V^2_{\star}(R)+V^2_{gas}(R)+V^2_{halo}(R)\right]$. The problem arise for the luminous components, because in general the stellar 
mass is unknown and a guess of the mass-to-luminosity ratio (M/L) is required. In contrast, the gas amount is usually well-known and rather small 
\citep{Swaters2000}. During the fitting procedure, generally the M/L is a free and constant parameter given the ignorance about the stellar content of a disk galaxy. 

{ The maximum disk hypothesis (MDH) was introduced by \citet{vanAlbada1985} and \citet{vanAlbada1986} to study the DM distribution of NGC 3198 and it is based on the 
assumption that the DM halo dominates at radii greater than 2.2 $R_D$. Inside this radius the luminous matter could account for the galaxy circular velocity estimated up 
to that radius. \citet{Ratnam2000} analysed the mass distribution of 83 spiral galaxies, using high resolution RCs data, formulated a more modern version of the MDH, 
suggesting a baryonic scale range in radius from $0.5\, R_D\, \sim\,$ 0.5 kpc to $2.0\, R_D\, \sim\,$ 30.0 kpc, where the luminous matter seems to produce a
substantial part of the gravitational potential of the studied galaxies.} 

{ In general there are two main factors that govern the degeneracy of the results of the RCs fitting: actual errors in RCs measurements and our ignorance of the true
distribution of luminous matter. Hypothetically, fitting different mass components to observed RCs should be a valid approach if we have a great number of high resolution data
with negligible measurement errors. In practice the observed RCs are far from these ideal conditions, therefore the RCs decomposition continue to be a degenerate problem. A more
robust determination of the distribution of the luminous matter based on more accurate constraints of the M/L variation with galactocentric radius and stellar colours could lessen
the degree of degeneracy.}

Some authors found a strong correlation between the variation of stellar M/L in optical and near-infrared (NIR) passbands and stellar population colours 
\citep{Bell2001, Bell2003}. \citet{Bell2001} also discovered that low luminosity galaxies have lower stellar M/L than high surface brightness galaxies and galaxies with few 
gas content.
From these works it remains clear that to remove the above mentioned degeneracy, due to our poor knowledge of the stellar M/L, more information about
the radial variation of the stellar M/L and the correlation with other galaxies properties are necessary. \citet{Gallazzi2005} explore the dependency
among the total galaxy stellar mass, the stellar metallicity, and the stellar age, estimating these parameters for a magnitude-limited sample of $\sim$ 
200,000 galaxies selected from SDSS DR2. These authors made a comparison of the observed galaxy spectra and a large library of synthetic spectra based on Monte 
Carlo library of star formation histories (SFHs). \citet{Gallazzi2005} and \citet{Gallazzi2008} found a large scatter in the relations amidst stellar mass, age 
and metallicity, concluding that the physical parameters of galaxies at the present time are not uniquely determined by stellar masses, and that gas infall or 
outflow could also play an important role in building galactic physical properties. 

From these works it is evident that the stellar M/L rely on so many physical parameters that the simplistic assumption of considering the stellar M/L constant 
along the galactic disk is unreal. In this respect, stellar population synthesis models (SPs) represent a valuable tool to constrain in a more realistic manner 
the true value of the stellar M/L. \citet{Zibetti2009} conceive a novel 
approach to build spatially resolved maps of stellar mass surface density considering broad-band photometry images of galaxies in the optical and 
NIR. The authors used a Monte Carlo library of 50,000 synthetic SPs by \citet{Bruzual2003} and \citet{Bruzual2007} incorporating an updated recipe for the 
thermally pulsing asymptotic giant branch stellar evolutionary phase following \citet{Marigo2007} and \citet{Marigo2008}. \citet{Gallazzi2009} studied 
the accuracy in the determination of stellar M/L from galaxy colours, using the SPs models of \citet{Bruzual2003} and a Monte Carlo library of 150,000 
SFHs, to examine which spectroscopic conditions and spectral signal-to-noise ratios (S/N) are required to decrease the errors on M/L determination for 
different galaxy types. The authors concluded that in galaxies with old stellar populations the accuracy of M/L is beneath 0.05 dex for S/N $>$ 20, 
while for galaxies with young stellar population the M/L presents, in general, higher statistical uncertainties. The results of \citet{Gallazzi2009} 
are based on the very strong assumption that the dust extinction is negligible. \citet{MacArthur2005} prove that the effect of dust attenuation is 
important near 4000 $\AA$, additionally, from the literature it is well-known that dust effects have a crucial impact on M/L values based on 
colours \citep{Zibetti2009}. For this reason, at the moment, it does not exist a robust analysis discussing the feasibility of SPs studies to determine 
the M/L, nonetheless SPs seem to be the more powerful tool to obtain the stellar mass content of galaxies.\\ 
The $\Lambda$-CDM paradigm \citep{Blumenthal1984} predicted cuspy DM density profiles in the centre of galaxies \citep{Navarro1996} and recent 
results with higher resolution seem to suggest a shallower DM density profile \citep{Navarro2004, Navarro2010}. Nonetheless the past and newer 
developments are in contradiction with observational results of cored DM density profiles in dwarf galaxies \citep{deBlok2001, deBlok2002, 
Swaters2003, Chemin2011}. This discrepancy is known as the cuspy-core problem. Many other authors, in an attempt to contribute to the solution of the 
cuspy-core problem, performed the RC decomposition considering a constant M/L \citep{Carignan1985, Carignan1988, Jobin1990, Persic1990, Martimbeau1994, 
Persic1996, Blais-Ouellette2001, deBlok2001, deBlok2002, Swaters2003, KuziodeNaray2006, Fuentes-Carrera2007, Spano2008, KuziodeNaray2008, Repetto2010, 
Chemin2011} and reinforce the evidence of a cored DM distribution in the central part of galaxies.    
Conversely, other authors pursue { spectrophotometric and} SPs studies to derive the M/L of the stellar component \citep{Salucci2008, Repetto2013, 
deDenus-Baillargeon2013} or at least to constrain the stellar component \citep{Dutton2005} to avoid the disk-halo degeneracy. 
{ \citet{Salucci2008} accomplished RCs mass modelling and spectral energy distribution fitting with spectrophotometric models to obtain the disk masses 
of 18 Sa spiral galaxies (principally bulgeless systems) finding that by decomposing the RCs with the spectrophotometric disk masses 
the results are consistent with the corresponding maximum disk solution.} \citet{Repetto2013} employed the \citet{Zibetti2009} 
method to optical (SDSS) and NIR (2MASS) images of NGC 5278 (KPG 390A) to obtain the stellar disk mass profile of that galaxy from SPs models to 
reproduce the RC of KPG 390A. The new strategy relied on fitting only the DM RC, obtained by subtracting the SPs baryonic disk from the observed RC of 
NGC 5278. The most important finding of \citet{Repetto2013} is that the favoured DM distribution is cored when the disk mass approximate to the 
maximum disk solution in agreement with the general belief. \citet{deDenus-Baillargeon2013} used a chemo-spectrophotometric galactic evolution model to 
determine the stellar M/L and perform the RC decomposition of ten spiral and dwarf irregular galaxies from the SINGS survey \citep{Kennicut2003}. The 
authors employed the settled baryonic disk as a weighting function to fit the model DM halo to the RCs of the studied subsample of galaxies, 
and concluded that the stellar disks obtained from their chemo-spectrophotometric models were compatible with the maximum disk 
hypothesis. The few examples presented, indicate that a growing effort to break the disk-halo degeneracy exists, however the most 
significant contributions trying to address the cuspy-core problem still rely on the general assumption of considering the stellar M/L constant along 
the galactic disk. In general, it is still missing a significant endeavour to earn the DM distribution, through RC fitting, determining the disk 
stellar mass from SPs studies. For this reason it is worth considering the cuspy-core problem with a different observational approach, focusing on
the formulation of a general procedure to better constrain the baryonic disk mass in galaxies.\\
This is the first of a series of papers dedicated to the study of the outer and inner mass distribution in irregular galaxies, applying and testing 
the novel method of \citet{Repetto2013}, using the RCs data of the WHISP HI survey of galaxies \citep{Verheijen2001}, to investigate the outer mass 
distribution, and the GHASP H$\alpha$ survey of galaxies, to analyse the inner mass distribution \citep{Amram2002, Garrido2002, 
Spano2008, Epinat2008a}. In this first article we concentrate on the determination of the mass distribution of two irregular galaxies UGC 4284 
and UGC 11861, whose RCs are somewhat symmetric and regular and for this reason more suitable to perform the RC analysis. 
The novelty of the approach conceived in \citet{Repetto2013} resides in deriving the stellar disk mass of the galaxy from
SPs studies and in subtracting the built baryonic disk from the observed RC to fit only the remaining DM RC as we have already summarised above.\\ 
The main motivation of these works are attempt to answer the following question: Performing the RC decomposition with more realistic M/L obtained from SPs 
studies, could change the determination of the DM halo parameters substantially? In the case of a positive answer, this research could shed new 
light on how the DM halo parameters change because of a non parametric baryonic component. In the case of a negative answer the analysis 
is still worth to characterise a novel procedure to constrain the stellar disk parameters with a deeper knowledge of the physics of the underlying 
stellar disk populations derived from photometric data.

This paper is organized as follows. Sections 2 and 3 describe the GALFIT disk/bulge/bar decomposition.
Section 4 describes the constrains of the baryonic M/L ratio through SPs.
Section 5 describes the H$\alpha$ and HI RC derivations.
Section 6 presents the analysis of the inner and outer DM distributions.
Section 7 discusses the results. Finally, Section 8 presents our conclusions. 

\section{2D GALFIT Disk/Bulge/Bar decomposition of UGC 4284}~\label{gal_dec0}

UGC 4284 is classified as Scd(s) in the RC3 \citep{deVaucouleurs1991}, and as SABc by \citet{Paturel2003}. 
\citet{Kandalian1998}, \citet{Chapelon1999} and \citet{CabreraLavers2004} found evidence of a central isophotal distortion that they explain postulating 
the existence of a bar with diameter of 38$\arcsec$. Other authors did not confirm the presence of a bar in UGC 4284, nevertheless they encountered 
non-circular and off-plane gas motion in the inner region of UGC 4284 that they discuss as originated by a warped gaseous disk \citep{Coccato2004, Jozsa2007}. 
Bearing in mind the purpose of the present study it is important to determine which kind of baryonic components do actually exist for UGC 4284, and given the 
incongruity of the above reported results we required to perform 2D isophotal analysis of the surface brightness of this galaxy, using GALFIT 
\citep{Peng2002, Peng2010} and considering at least three principal components: disk, bulge and bar. 
UGC 4284 has been observed with Hubble Space Telescope (HST) in the filters F555W (WFPC2 V), F814W (WFPC2 wide I) and F606W 
(WFPC2 wide V), nonetheless the filters F555W and F814W show only a small part of the periphery of the galaxy. In contrast the filter F606W exhibits 
a larger portion of the disk of UGC 4284, so we use that filter to decompose the disk of UGC 4284. The results of the 2D exponential disk fit are 
presented in Table~1. In the case of UGC 4284 we also examined the possibility of a bulge and a bar components given the inconsistency in the morphological 
classification of this galaxy. We use the Spitzer IRAC 4.5 $\mu$m image because UGC 4284 is indiscernible in the 2MASS J, H, K bands, and in all the SDSS 
filters is too faint to allow an obvious separation from the background. Neither the Sersic component nor the Ferrers \citep{Ferrers1877} component produce 
acceptable fit parameters to reveal the presence of a bulge and a bar, for this reason we conclude that UGC 4284 has only an exponential disk component in 
total agreement with the RC3 classification. 

\setcounter{table}{0}
\begin{table}
\caption{Results of GALFIT isophotal analysis for UGC 4284.}
\begin{tabular}{@{}lccc}
\hline
UGC 4284 exponential disk final parameters.\\
\hline
Surface Brightness (mag arcsec$^{-2}$) & 17.97\\
Scale length (kpc) & 1.27\\ 
Axial ratio & 0.46\\
Inclination (degrees) & 64.98\\
Position Angle (degrees) & 44.45\\
\hline
\end{tabular}
\end{table}

\section{2D GALFIT Disk/Bulge/Bar decomposition of UGC 11861}~\label{gal_dec1}

UGC 11861 was classified as SABdm in the RC3 \citep{deVaucouleurs1991}, as SABd by \citet{Paturel2003}. At first glance the presence of a 
bar is not so evident from single optical (SDSS) and NIR (2MASS) broad band images, nevertheless a central boxy structure is clearly visible
in the Spitzer IRAC 3.6 $\mu$m image. The congruity in the morphological taxonomy of UGC 11861 allow us to identify promptly 
at least two components for that galaxy: a disk and a bar. We use GALFIT to prove the existence of such structures for UGC 11861. We employ 
the SDSS g band image for the disk and the Spitzer IRAC 3.6 $\mu$m image for the bar. We consider an exponential disk and Ferrers model 
\citep{Ferrers1877} for the bar. The GALFIT result for the disk and bar GALFIT model are listed in Table~2. We also consider 
the possibility of a bulge component for UGC 11861, for which we adopt a Sersic profile and we fit with GALFIT 2MASS J, H, and K band images 
regarding a de Vaucouleurs bulge with an older stellar population, and SDSS g,r and i band images considering the possible presence of pseudobulges 
with younger stellar population. The choice of that photometric range to examine the possible existence of a bulge for UGC 11861 is also dictated 
by the fact that the bar emission is fainter in the selected passbands, and this fact could in principle allow an unambiguous detection of a potential 
bulge. In the case of the SDSS g, r and i bands we first fit an exponential disk model and then we try to fit a Sersic component to detect the existence of 
an hypothetical bulge in the residuals without success. The GALFIT fit to the 2MASS J, H, and K passbands did not reveal the presence of a bulge. From the 
GALFIT analysis we conclude that UGC 11861 is composed of an exponential disk and a bar, in good agreement with the morphological classification of that 
galaxy. 

\setcounter{table}{1}
\begin{table}
\caption{Results of GALFIT isophotal analysis for UGC 11861.}
\begin{tabular}{@{}lccc}
\hline
UGC 11861 exponential disk final parameters.\\
\hline
Surface Brightness (mag arcsec$^{-2}$) & 19.35\\
Scale length (kpc) & 1.30\\ 
Axial ratio & 0.47\\
Inclination (degrees) & 64.27\\
Position Angle (degrees) & 71.46\\
\hline
UGC 11861 Ferrers bar final parameters.\\
\hline
Surface Brightness (mag arcsec$^{-2}$) & 8.83\\
Outer truncation radius (kpc) & 2.48\\ 
$\alpha$ & 2.95\\
$\beta$  & 0.0\\
Axial ratio & 0.38\\
Inclination (degrees) & 70.74\\
Position Angle (degrees) & -16.6\\
\hline
\end{tabular}
\end{table}

\section{Baryonic $M/L$ ratio from resolved mass maps}~\label{mass_maps}

Reliable $M/L$ ratios are an important aspect when
determining the baryonic contribution to the rotation curves of disk galaxies.
By using a chemo-spectrophotometric galactic evolution model~\citet{deDenus-Baillargeon2013}
highlighted the importance of adopting a radially varying $M/L$ ratio, vs. a radially constant one~\citep[see also][]{Repetto2013}.

In this work we adopt a resolved mass map method~\citep{Zibetti2009} that relies on a Monte Carlo
library of 50,000 stellar population spectra randomly drawn from the SPS models of S. Charlot \& G. Bruzual (CB07).
These models use single stellar populations as described by~\citet{Bruzual2003},
but include the prescription for stellar evolution in the thermally-pulsating
asymptotic giant branch (TP-AGB) by~\citet{Marigo2007}. 
The IMF of~\citet[Galaxy-disk,][]{Chabrier2003} is adopted.
The adjustable parameters of the models are the
star formation history (SFH), which is exponentially declining with random bursts superimposed,
the metallicity, uniformly distributed between 0.02 and 2 times solar,
and the extinction by dust, based on the two-component model of~\citet{Charlot2000}.\footnote{
The model takes into account the fact that stars are born in dense molecular clouds.
A different treatment is given for the attenuations of young and old stars.}
The Monte Carlo library is obtained for a given combination of colours,
and the mass-to-light ratio ($M/L$) for a given filter.

\begin{figure}
\centering
\includegraphics[width=1.0\hsize]{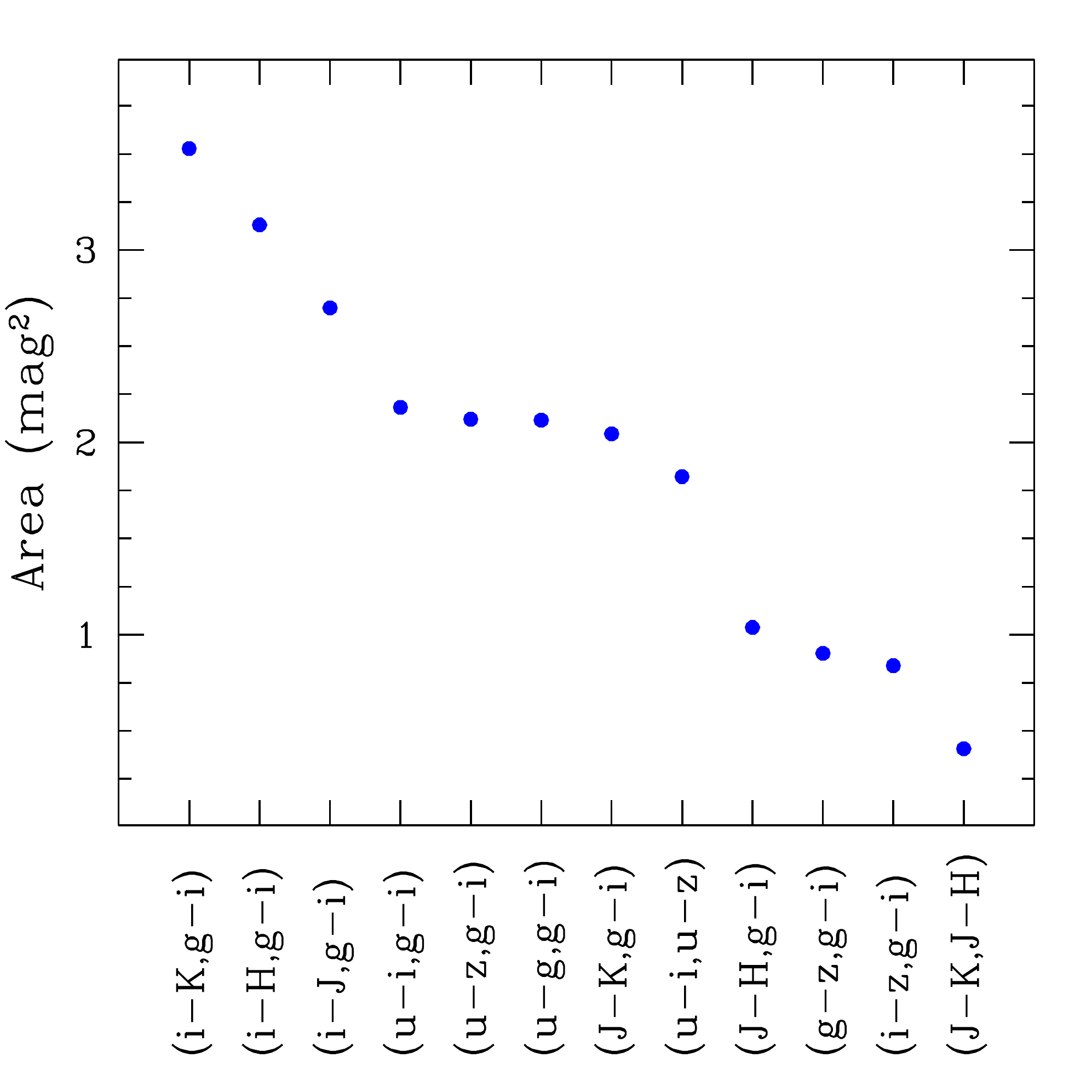}
\caption[f1.pdf]{Area covered in a colour-colour diagram,
for different colour combinations obtained from a Monte Carlo SPs library grouped
in colour bins of $0.05\times0.05$ mag$^2$, as described in~\citet{Zibetti2009}. Filters $u$, $g$, $i$, and $z$ from SDSS.
Filters $J$, $H$, and $K$ from 2MASS.
~\label{fig1}}
\end{figure}

~\citet{Zibetti2009} fiducial method involves the use of the $g$
and $i$ bands from SDSS, and one NIR filter.
The main reason for the use of this combination of filters is that it provides the less degeneration
for the SPs models in a colour-colour diagram (e.g., $g-i$, vs. $i-H$).
Other combinations of filters, e.g, Johnson's $V$ and $I$ combined with one NIR filter, 
show a smaller area in the colour-colour diagrams, thus providing 
more degeneracy, i.e., the same colours can be attributed to many different model
parameters. The same applies to various other combinations of filters~(see figure~\ref{fig1}).
It was found that the absolute mass values are overestimated, as compared to the fiducial method,
when a combination of colours with more degeneracy is adopted.

The $M/L$ is obtained for the NIR filter as a function of $(g-i)$ and, e.g., $(i-H)$.
The models are grouped in bins that take into account the observational error in the
photometry, typically 0.05 mag. For each bin, the median value of $M/L$ is computed.
The final look-up tables can be compared with observations
in a pixel-by-pixel basis. A map of $M/L$ is obtained which can be converted to absolute mass
if the distance to the object is known.

\begin{figure*}
\centering
\includegraphics[width=1.0\hsize]{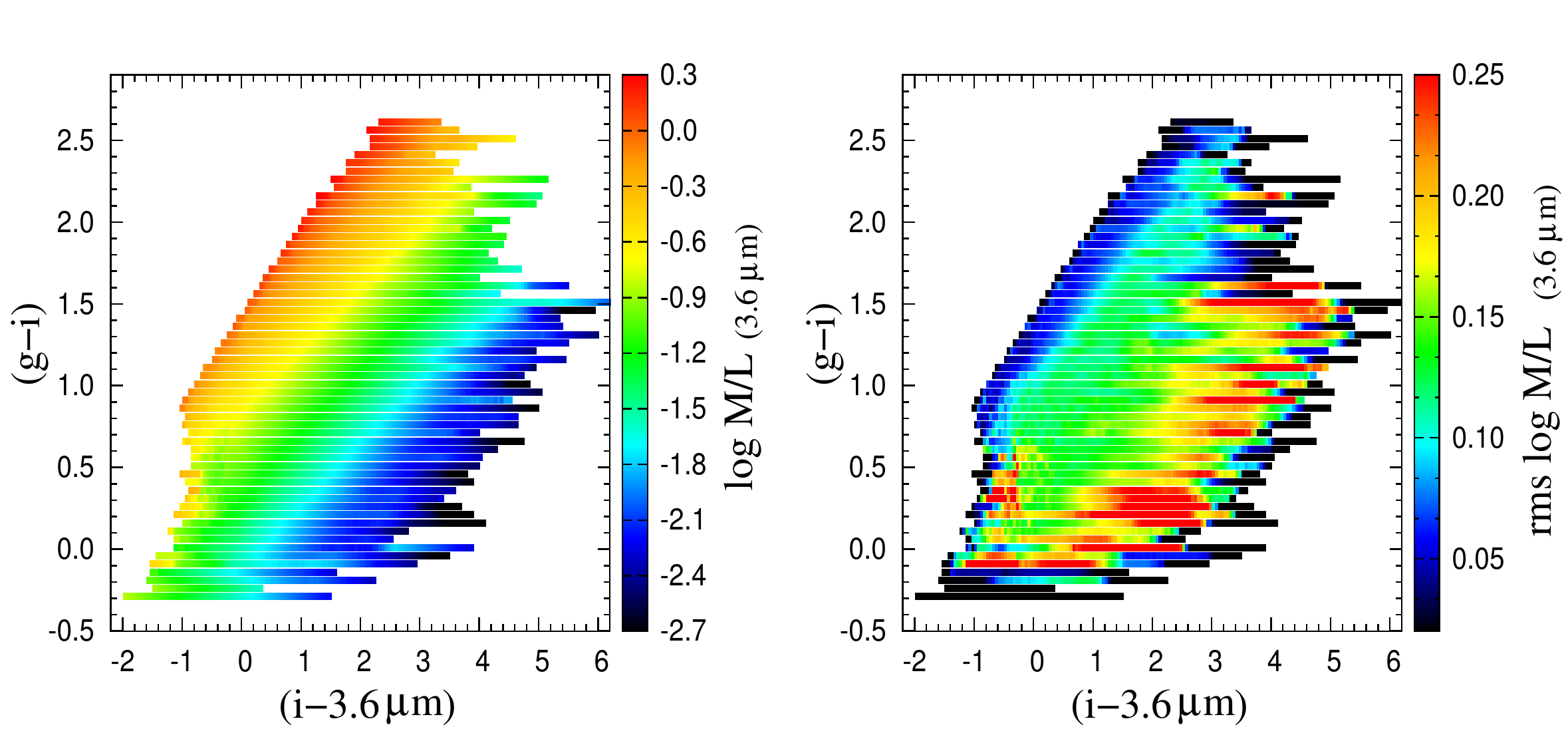}
\caption[f2.pdf]{ {\it{Left}}: Logarithm of the effective, i.e., as seen by the observer,
$M/L$ ratio for the $3.6\micron$ filter, $M/L_{(3.6\micron)}$,
in a $(g-i)$, vs., $(i-3.6\micron)$, colour-colour diagram. Data are grouped in bins of $0.05\times0.05$ mag$^2$.
{\it{Right}}: Logarithm of the root mean square (rms) errors, corresponding to the $M/L_{(3.6\micron)}$ data in the left panel.
The mean error value (of all plotted data) is~$\sim30\%$.
~\label{fig2}}
\end{figure*}

When analysing the spectral energy distribution (SED) of galaxies, the wavelength range
between $2.5\micron$ and $10\micron$ is considered to be a mixture of stellar
and dust emission~\citep[e.g.,][]{daCunha2008}.
In order to apply the resolved mass map method~\citep{Zibetti2009} to get a $M/L$ map of,
e.g., the IRAC $3.6\micron$ filter, an infrared SED model is needed.
The model of~\citet{daCunha2008} is adequate for this purpose.
This model incorporates four main dust components,
the emission from polycyclic aromatic hydrocarbons (PAHs), the mid-IR continuum from hot dust,
and both the emission from warm and cold dust in thermal equilibrium.
An energy balance condition is applied in a way that
the total energy absorbed by dust, in stellar birth clouds and in the diffuse ISM, is 
re-radiated in wavelengths over the range from 3$\micron$ to 1000$\micron$.
It is assumed that the dust in the ISM is heated only by starlight.
The infrared model can be easily combined with the SPs models of~\citet{Bruzual2003} and CB07.
A Monte Carlo library of 50,000 dust emission spectra, for wide ranges of dust temperatures
and fractional contributions by the different dust components, is provided in
the MAGPHYS\footnote{http://www.iap.fr/magphys/magphys/MAGPHYS.html} package of~\citet{daCunha2008}.
We combine this library with the SPs library
taking care of the energy balance condition as described in~\citet{daCunha2008}. A total
of 667 million spectra were obtained. The models were grouped in bins of $0.05\times0.05$ mag$^2$,
for the $(g-i)$, and $(i-3.6\micron)$ colours. The mean $M/L$ ratio for the $3.6\micron$ filter
was obtained for each bin. The results are shown in figure~\ref{fig2}, together with the rms values.

\begin{figure*}
\centering
\includegraphics[width=1.0\hsize]{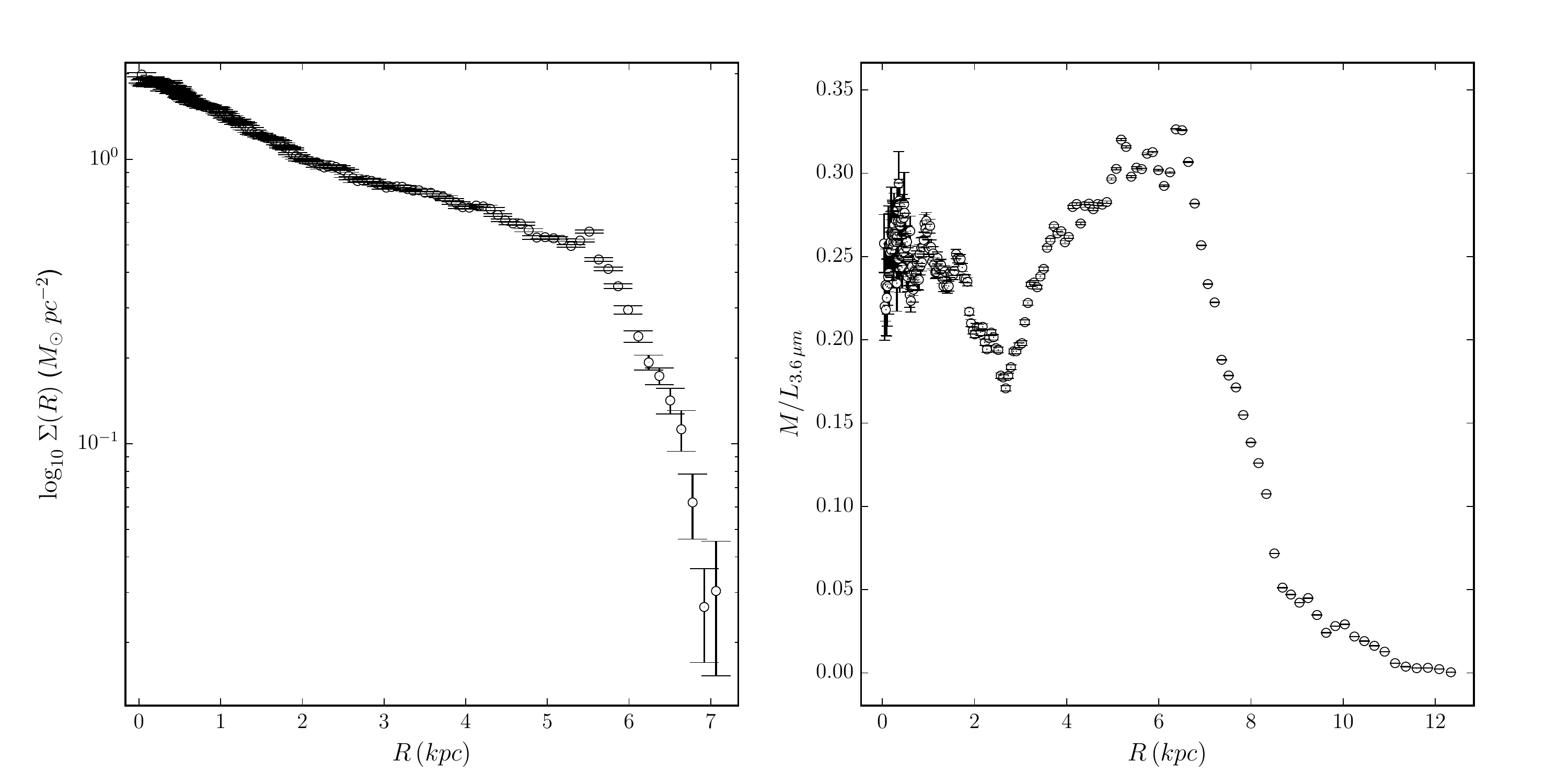}
\caption[f3.pdf]{
Radial profiles for UGC~4284.
{\it{Left}}: Azimuthal average of the stellar mass surface density, $\mu_{*}$, as a function
of radius.
{\it{Right}}: Azimuthal average of the mass-to-light ratio, $M/L_{3.6\micron}$, as a function
of radius.
An inclination angle of $59\degr$, and position angle of $169.9\degr$, were adopted as deprojection parameters.
~\label{fig3}}
\end{figure*}

We then use these results (in figure~\ref{fig2}) to obtain the two-dimensional stellar mass map, for UGC~11861.
In figure~\ref{fig3} we show the azimuthally-averaged radial profiles of
the $M/L_{(3.6\micron)}$ ratio (left plot), and the stellar ($\mu_{*}$) mass surface density (right plot).
We assume an object distance of 18.4 Mpc, and correct the models for Galactic extinction~\citep{Schlafly2011,Chapman2009}.

\begin{figure*}
\centering
\includegraphics[width=1.0\hsize]{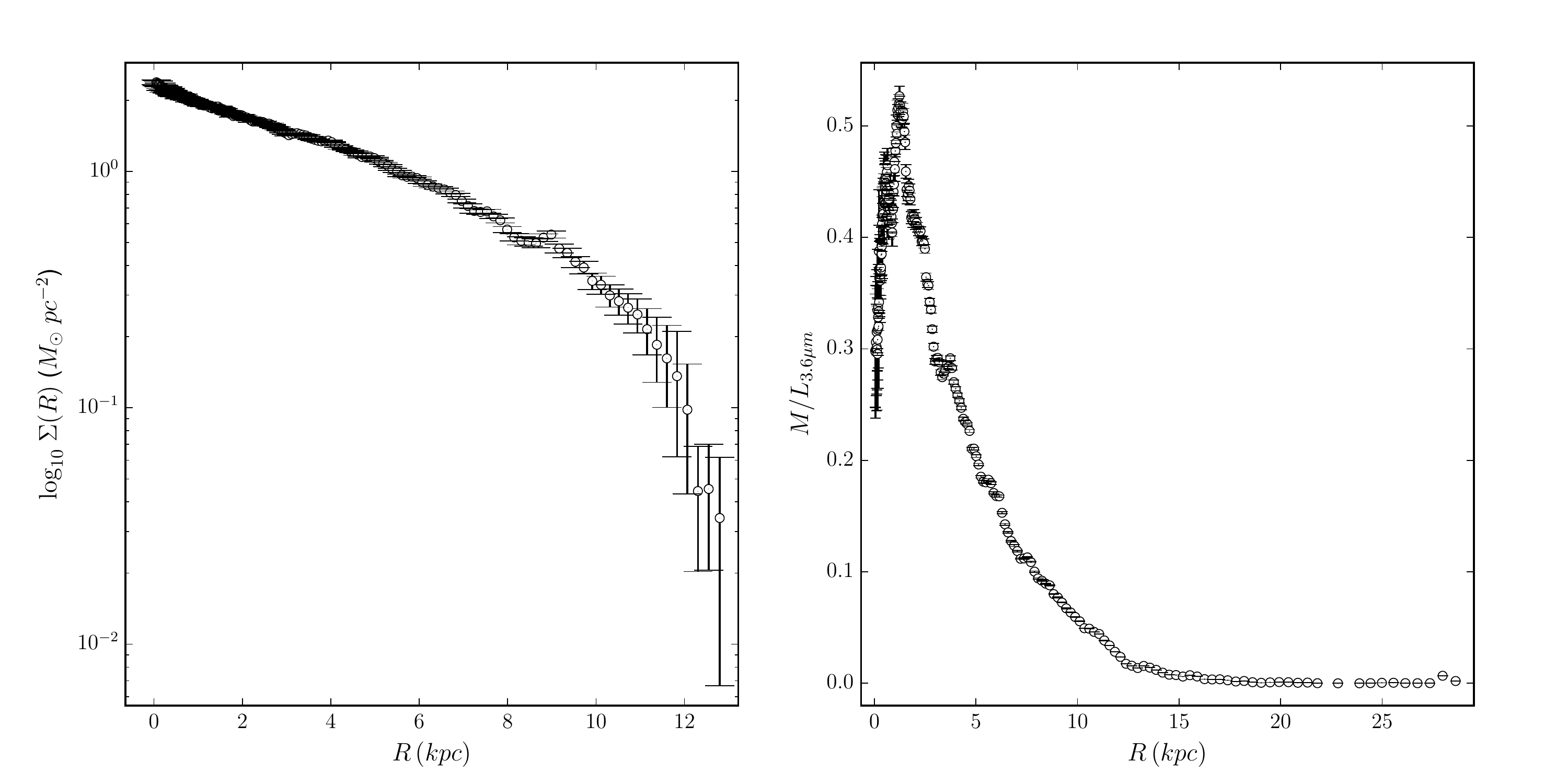}
\caption[f4.pdf]{
Radial profiles for UGC~11861.
{\it{Left}}: Azimuthal average of the stellar mass surface density, $\mu_{*}$, as a function
of radius.
{\it{Right}}: Azimuthal average of the mass-to-light ratio, $M/L_{(3.6\micron)}$, as a function
of radius.
An inclination angle of $61\degr$, and position angle of $10.7\degr$, were adopted as deprojection parameters.
~\label{fig4}}
\end{figure*}

For the case of UGC~4284 the resolved mass map method was also adopted. 
The $3.6\micron$ data was taken from the Spitzer Survey of Stellar Structures in galaxies \citep{Sheth2010}.
The resulting azimuthally averaged radial profiles for the $M/L_{3.6\micron}$ ratio, and the stellar mass
surface density ($\mu_{*}$) are shown in figure~\ref{fig4}. We assume a distance of 9.42 Mpc.

\subsection{Baryonic HI$+$He gas components of UGC 4284 and UGC 11861}~\label{gas_rcs}

\begin{figure*}
\centering
\includegraphics[width=1.0\hsize]{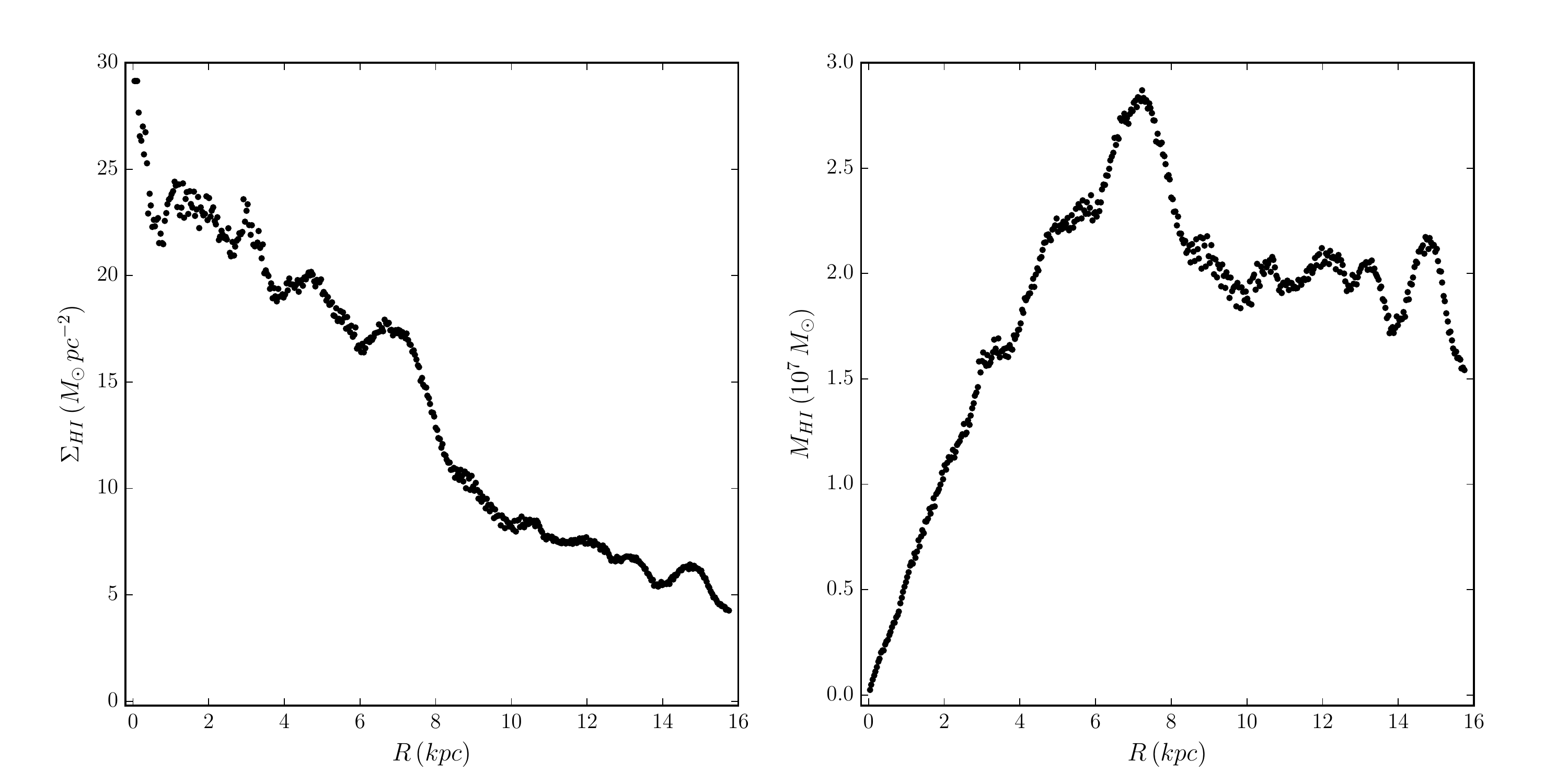}
\caption[f5.pdf]{
HI mass gas profiles for UGC~4284.
{\it{Left}}: Azimuthal average of the HI mass surface density, $\Sigma_{HI}$, as a function
of radius.
{\it{Right}}: Azimuthal average of the HI mass, $M_{HI}\,(10^7\,M_{\odot})$, as a function
of radius.
An inclination angle of $59\degr$, and position angle of $169.9\degr$, were adopted as deprojection parameters.
~\label{fig5}}
\end{figure*}

\begin{figure*}
\centering
\includegraphics[width=1.0\hsize]{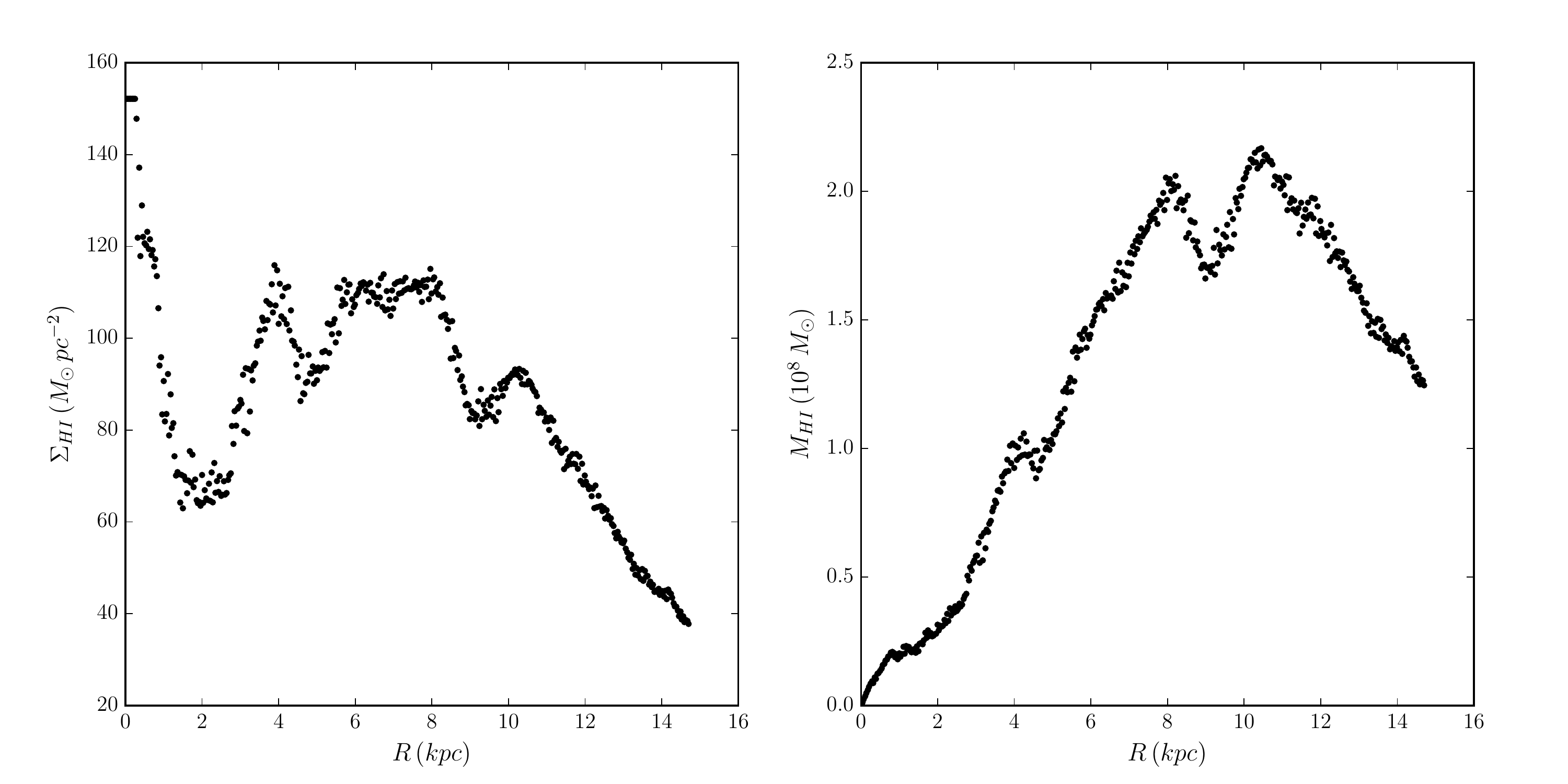}
\caption[f6.pdf]{
HI mass gas profiles for UGC~11861.
{\it{Left}}: Azimuthal average of the HI mass surface density, $\Sigma_{HI}$, as a function
of radius.
{\it{Right}}: Azimuthal average of the HI mass, $M_{HI}\,(10^8\,M_{\odot})$, as a function
of radius.
An inclination angle of $61\degr$, and position angle of $10.7\degr$, were adopted as deprojection parameters.
~\label{fig6}}
\end{figure*}

{ The HI gaseous RCs of both galaxies were obtained integrating the total HI column density map in elliptical rings using the task {\it ELLINT} in the
{\it GIPSY} package. The outputs are the HI mass surface density and the HI cumulative mass displayed in figures~\ref{fig5} and ~\ref{fig6}. The contribution of He
and metals was taken into account scaling by a factor of 1.4 the HI surface density and mass. The total HI$+$He RCs of both galaxies are displayed in 
figure~\ref{fig7}.}

\section{H$\alpha$ and HI RC of UGC 4284 and UGC 11861}~\label{hia_rcs}

\begin{figure*}
\centering
\includegraphics[scale=0.4]{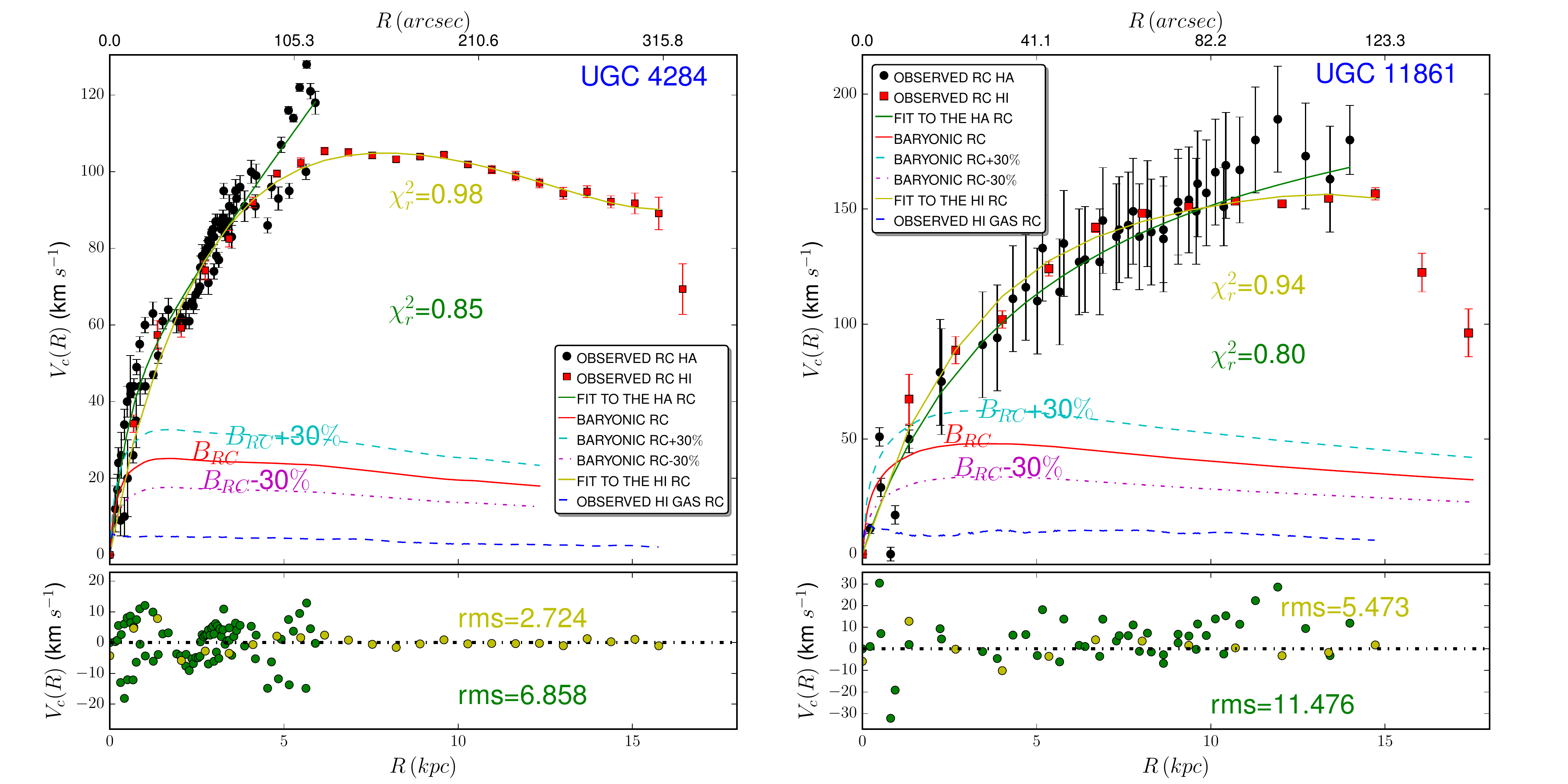}
\caption[f5.pdf]{
H$\alpha$ and HI RCs of UGC 4284 and UGC 11861 together with the corresponding baryonic 
stellar and gaseous (HI$+$He) RCs. The comparison between the two tracers is to motivate our choice to use the H$\alpha$ 
RCs to analyse the inner DM distribution and the HI RCs to estimate the outer DM distribution.
In this figure is also shown the preliminary fit to the H$\alpha$ and HI RCs with the corresponding 
$\chi^2$ values and residuals. The functional forms used for the preliminary fit are exponential-like 
for the H$\alpha$ RCs of UGC 4284 and UGC 11861 and polynomials for the HI RCs of both galaxies. We used
those preliminary fits instead of the observed H$\alpha$ and HI RCs to perform the subsequent analysis.
~\label{fig7}}
\end{figure*}

The H$\alpha$ RCs of UGC 4284 and UGC 11861 were derived in the frame of the Gassendi H$\alpha$ survey of spirals
\citep[GHASP][]{Garrido2002}, dedicated to study the 2D H$\alpha$ velocity field and RCs of a sample of 203 spiral and irregular galaxies. 
\citet{Garrido2003, Garrido2004, Garrido2005} perform the first data reduction and analysis of 97 galaxies in the GHASP sample. \citet{Epinat2008a}
accomplished again the reduction process of the 97 galaxies more 108 new galaxies, using adaptive binning technique based on Voronoi 
tessellation in order to provide different weights for low and high signal-to-noise zones in the velocity fields. The final 
high resolution 2D velocity fields of the GHASP survey have a spectral accuracy of about 5 km s$^{-1}$ and a spatial resolution of $\sim$ 2$\arcsec$.
From those high resolution 2D velocity fields, \citet{Epinat2008a} constructed the RCs for the most galaxies of that survey. 
A part of the GHASP sample (130 galaxies) has been previously observed by the Westerbork survey of HI in spiral galaxies (WHISP) 
intended to investigate the 2D HI kinematic of 500 spiral and irregular galaxies \citep{Verheijen2001}. The typical spectral resolution of the WHISP 
sample of 2D HI velocity fields is about 5 km s$^{-1}$ with a spatial resolution of $\sim$ 12$\arcsec$.\\ 
\citet{vanEymeren2011} obtained the HI RCs of a sample of 70 spiral and irregular galaxies, also including UGC 4284 and UGC 11861, using the WHISP HI 
velocity fields by applying a tilted-ring analysis to derive accurately the kinematic parameters from the velocity fields, obtaining residuals of a few 
km s$^{-1}$. In the particular case of UGC 4284 and UGC 11861, the HI WHISP RCs are relatively more extended that the H$\alpha$ RCs, they are regular and 
in the inner regions they have a good accordance with the H$\alpha$ RC. We decide to use the H$\alpha$ RCs to explore the inner DM distribution and the HI 
RCs to study the outer DM distribution. The comparison between the H$\alpha$ and HI RCs is shown in figure~\ref{fig7}. In the same figure are also 
presented the baryonic RCs, considering errors giving a variation of more and less 30$\%$ of the disk mass for both galaxies. From figure~\ref{fig7} 
it is clear that stars and gas alone are not enough to account for the total mass of UGC 4284 and UGC 11861. The total disk mass of UGC 4284 is 
$M_d=7.92\times10^9 M_{\odot}$, the total disk mass of UGC 4284 more 30$\%$ is $M_d=1.03\times10^{10} M_{\odot}$ and the total disk mass of UGC 4284
less 30$\%$ is $M_d=5.55\times10^9 M_{\odot}$. The total disk mass of UGC 11861 is $M_d=5.38\times10^{10} M_{\odot}$, the total disk mass of UGC 11861 
more 30$\%$ is $M_d=6.99\times10^{10} M_{\odot}$ and the total disk mass of UGC 11861 less 30$\%$ is $M_d=3.76\times10^{10} M_{\odot}$. In the case of
UGC 11861 the disk mass is the sum of the disk and the bar component (see section \ref{gal_dec1}). In the next two sections we explore the possibility 
of DM, for UGC 4284 and UGC 11861, to explain the discrepancy between the total observed mass distribution and the mass content due to stars and gas.

\section{Inner and Outer DM distribution for UGC 4284 and UGC 11861}~\label{io_dm}

We investigate the inner and outer DM content of UGC 4284 and UGC 11861 accomplishing the H$\alpha$ and the HI DM RCs fitting for both galaxies, considering
possible errors in the stellar disk determination of plus and minus 30$\%$ for all the DM models examined in this study. { The HI DM RCs were derived by 
subtracting the stellar RCs and also the HI$+$He gaseous RCs from the observed HI RCs of both galaxies. 

We consider the DM halo of \citet{Navarro1996} (NFW), the Burkert (BK) \citet{Burkert1995} DM halo and the \citet{DiCintio2014a} DM halo (DC). The DC DM halo
was formulated from the double power-law model of \citet{Zhao1996} expressing the three power-law exponents as functional forms of the stellar-to-halo mass ratio, 
establishing a two free parameters DM halo. We use the prescription of \citet{Dutton2014} for the concentration parameter for the NFW DM halo.} 

In the present study we use the {\it minuit} minimizer through the ROOT package written in \Cplusplus. The ROOT package is a huge 
object-oriented data analysis framework developed at CERN that allows sophisticated analysis of practically any kind of data. The {\it minuit} minimizer uses the 
MIGRAD algorithm that implements a variable metric method, with inexact line search checking for positive-definiteness. The main difference with respect to the 
Levemberg-Marquardt method is that, the minimization algorithm only converges on the global minimum, for this purpose an important output parameter is the 
estimated distance from the (global) minimum (Edm), a quantity that informs the user when the global minimum is reached. The Edm together with the $\chi^2$ 
values and also the $\chi^2$ filled contours, represent the fundamental criteria used in this study to assess the reliance of the determined solutions. 

{ The fitting procedure employed in this study considers as free parameters the DM halos masses and radii respectively within the intervals [$10^{8}, 10^{14}$] $M_{\odot}$ 
and [0.1, 100.0] kpc. The solutions we have found are listed in tables~2, 3, 4 and 5. In addition, the results with stellar disk mass $M_d=7.92\, \times\, 10^9\, M_{\odot}$ for 
UGC 4284 and stellar disk mass $M_d=5.38\, \times\, 10^{10}\, M_{\odot}$ for UGC 11861 are displayed in figures~\ref{fig8}-\ref{fig13}. The $\chi^2$ filled contours of the 
solutions presented in figures~\ref{fig8} to ~\ref{fig13} are exhibited in figures~\ref{fig14} and ~\ref{fig15}.   

\begin{figure*}
\centering
\includegraphics[scale=0.4]{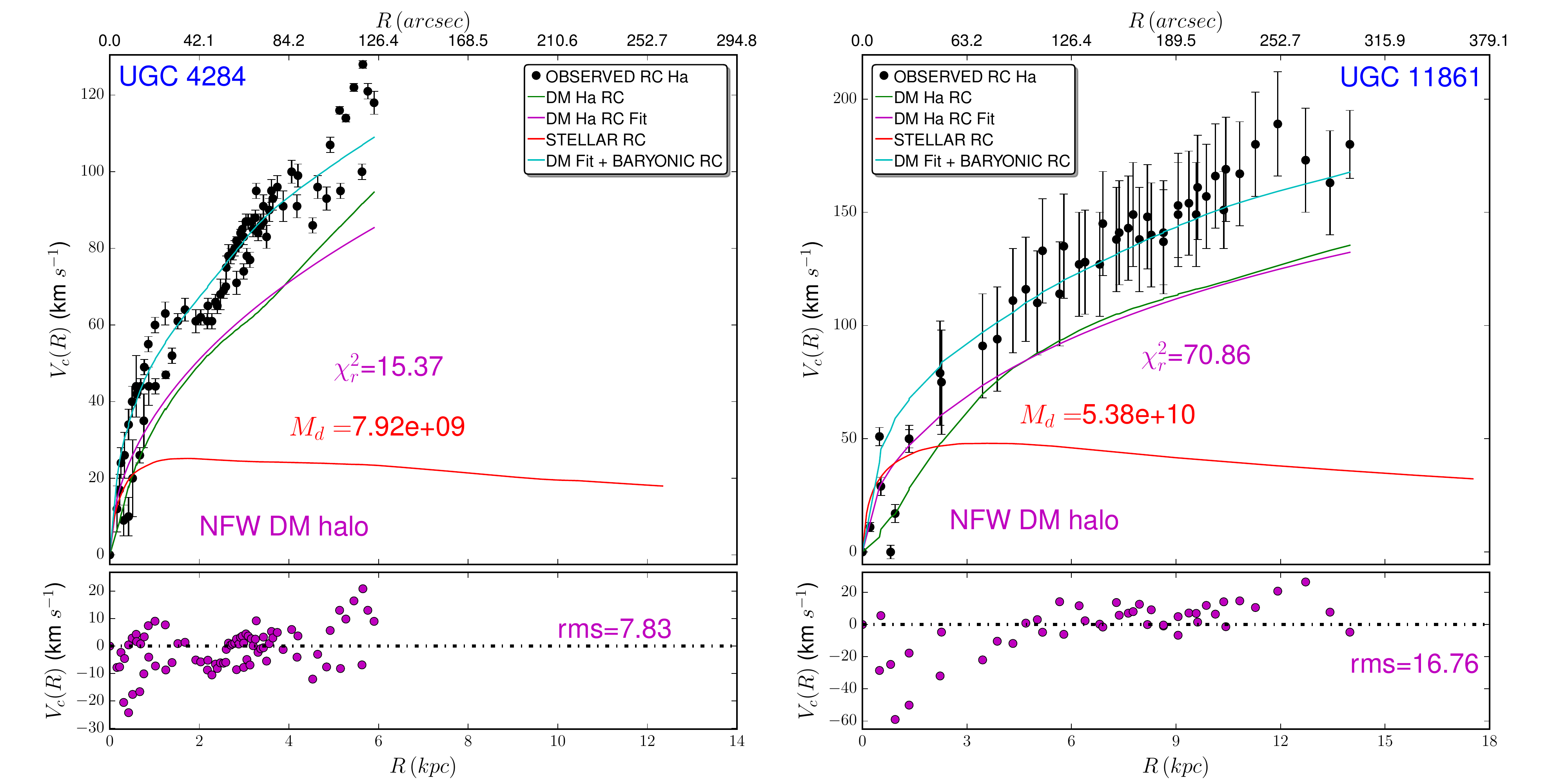}
\caption[f8.pdf]{{\it{Left}}: NFW fit of H$\alpha$ DM RC of UGC 4284. {\it{Right}}: NFW fit of H$\alpha$ DM RC of UGC 11861.
In both cases the $\chi^2$ contours contain less than four points. For this reason we do not show that information for these
solutions. The number of degrees of freedom (DOF) are 78 for UGC 4284 and 43 for UGC 11861.
~\label{fig8}}
\end{figure*}

\begin{figure*}
\centering
\includegraphics[scale=0.4]{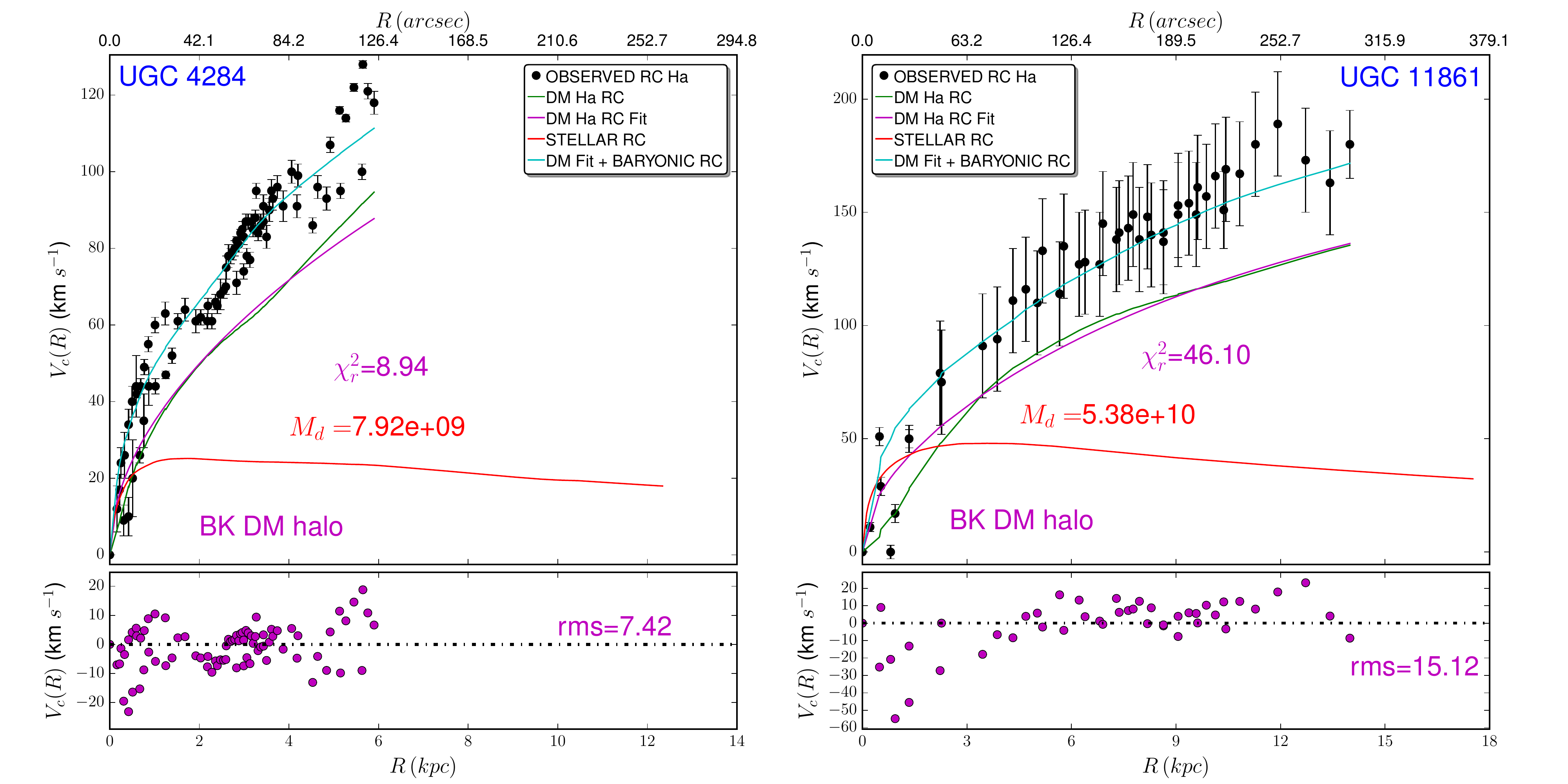}
\caption[f9.pdf]{{\it{Left}}: BKH fit of H$\alpha$ DM RC of UGC 4284. {\it{Right}}: BKH fit of H$\alpha$ DM RC of UGC 11861.
The number of degrees of freedom (DOF) are 78 for UGC 4284 and 43 for UGC 11861. The $\chi^2$ contours are displayed in 
figure~\ref{fig14}.
~\label{fig9}}
\end{figure*}

\begin{figure*}
\centering
\includegraphics[scale=0.4]{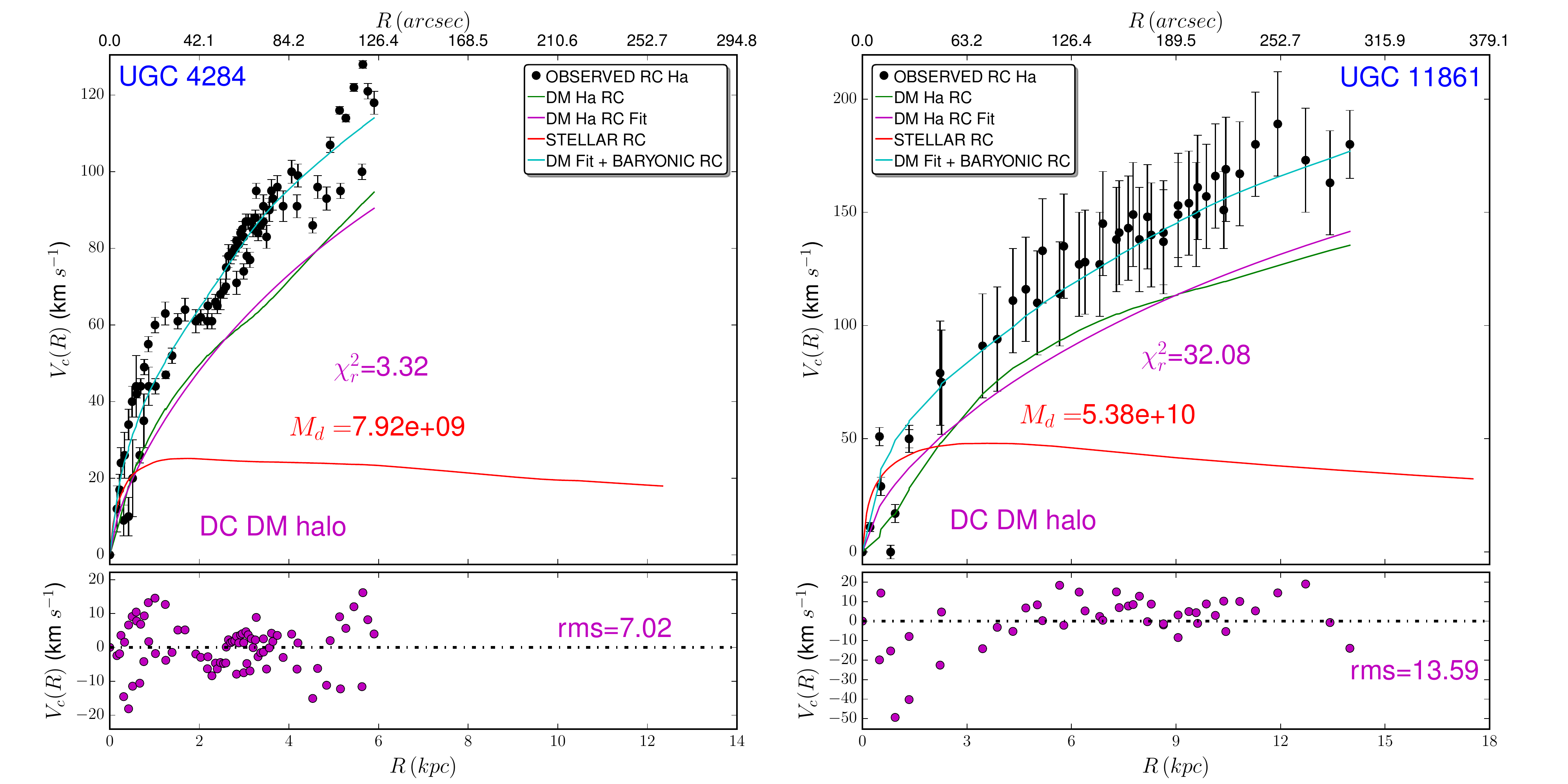}
\caption[f10.pdf]{{\it{Left}}: DCH fit of H$\alpha$ DM RC of UGC 4284. {\it{Right}}: DCH fit of H$\alpha$ DM RC of UGC 11861.
The number of degrees of freedom (DOF) are 78 for UGC 4284 and 43 for UGC 11861, the corresponding contours are displayed in 
figure~\ref{fig15}.
~\label{fig10}}
\end{figure*}

\begin{figure*}
\centering
\includegraphics[scale=0.4]{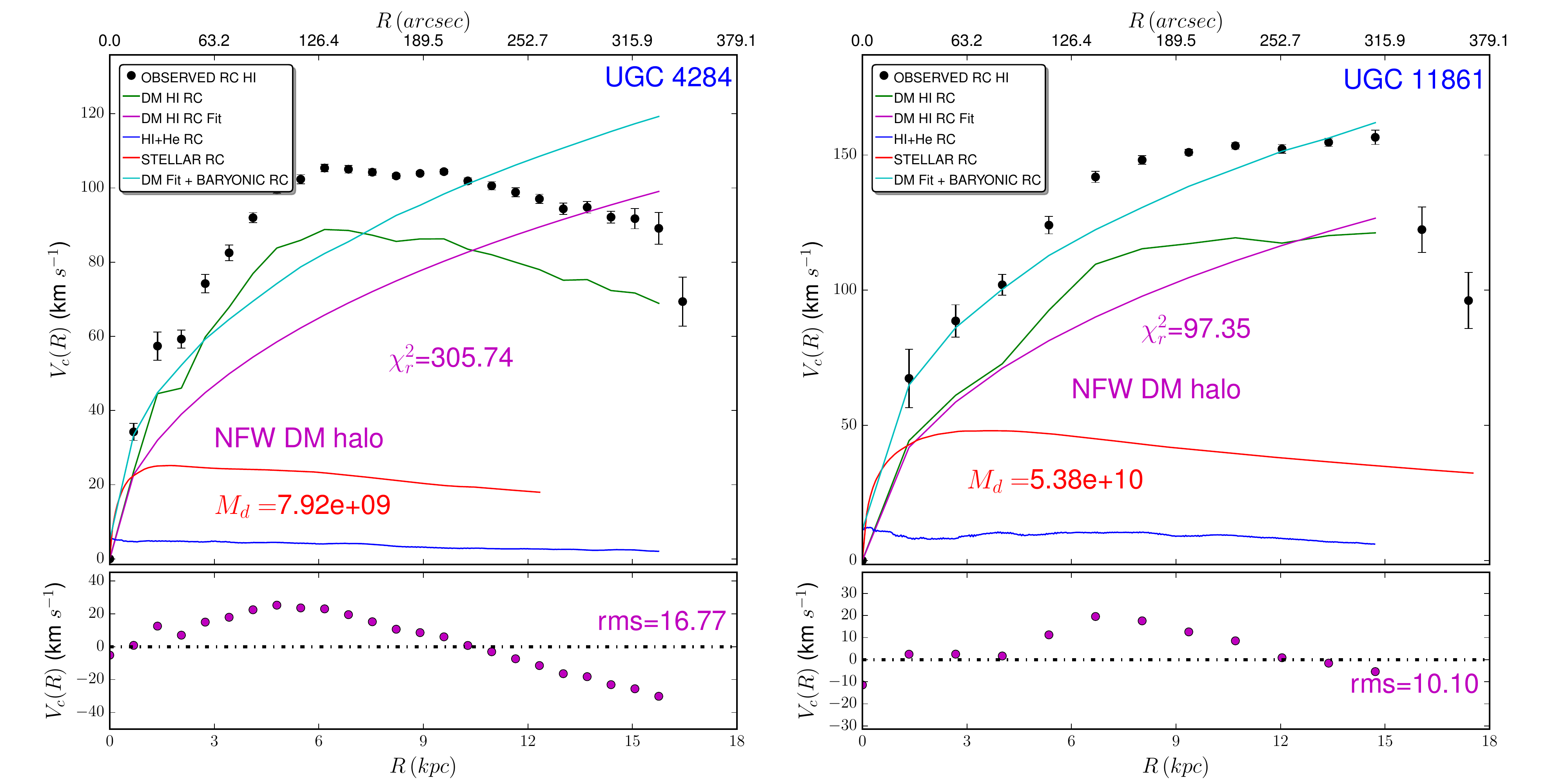}
\caption[f11.pdf]{{\it{Left}}: NFW fit of HI DM RC of UGC 4284. {\it{Right}}: NFW fit of HI DM RC of UGC 11861. In both cases 
the $\chi^2$ contours contain less than four points for this reason we do not show that information for these solutions.
The number of degrees of freedom (DOF) are 22 for UGC 4284 and 10 for UGC 11861.
~\label{fig11}}
\end{figure*}

\begin{figure*}
\centering
\includegraphics[scale=0.4]{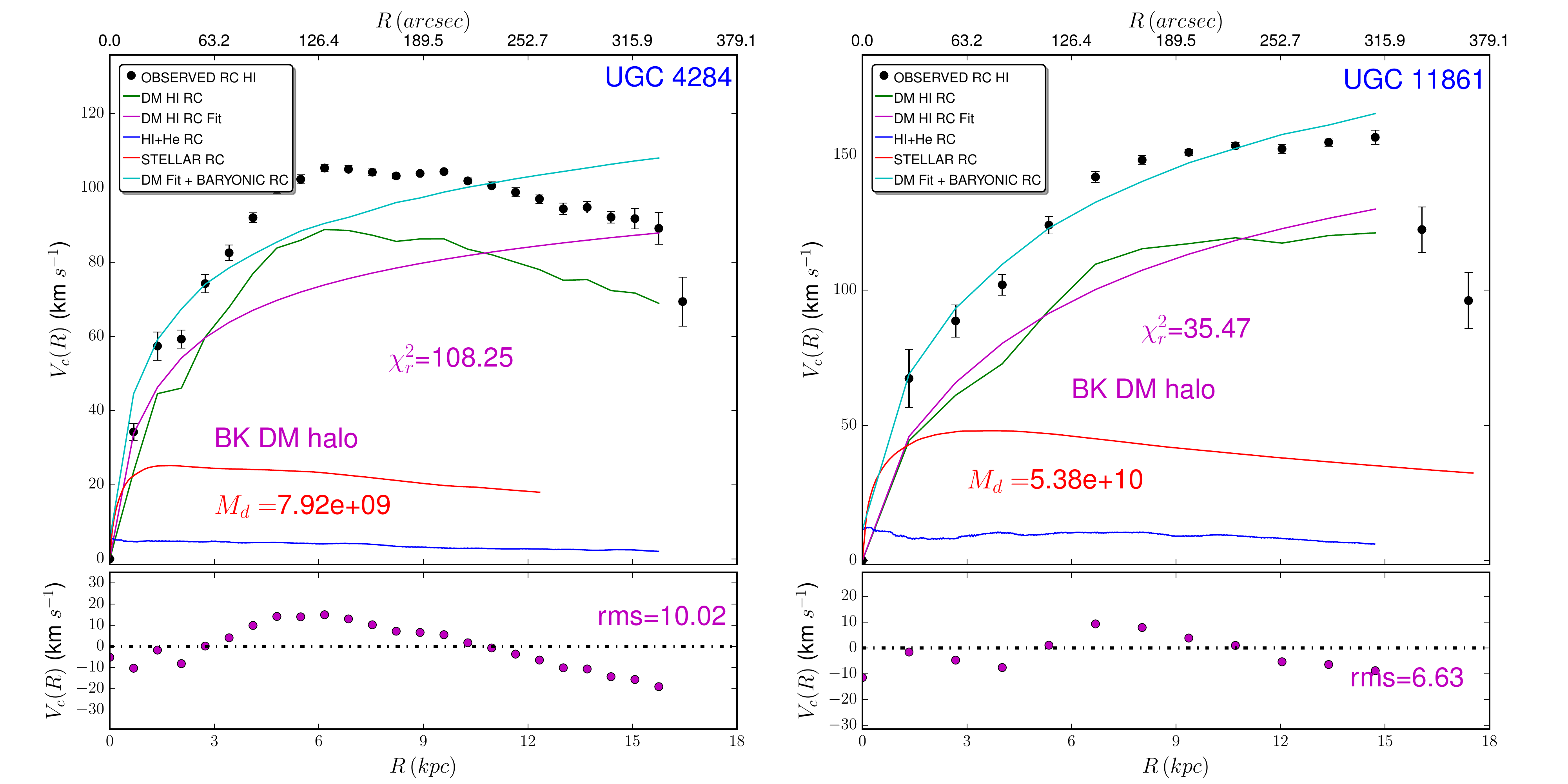}
\caption[f12.pdf]{{\it{Left}}: BKH fit of HI DM RC of UGC 4284. {\it{Right}}: BKH fit of HI DM RC of UGC 11861.
The number of degrees of freedom (DOF) are 22 for UGC 4284 and 10 for UGC 11861. The $\chi^2$ contours are displayed 
in figure~\ref{fig14}.
~\label{fig12}}
\end{figure*}

\begin{figure*}
\centering
\includegraphics[scale=0.4]{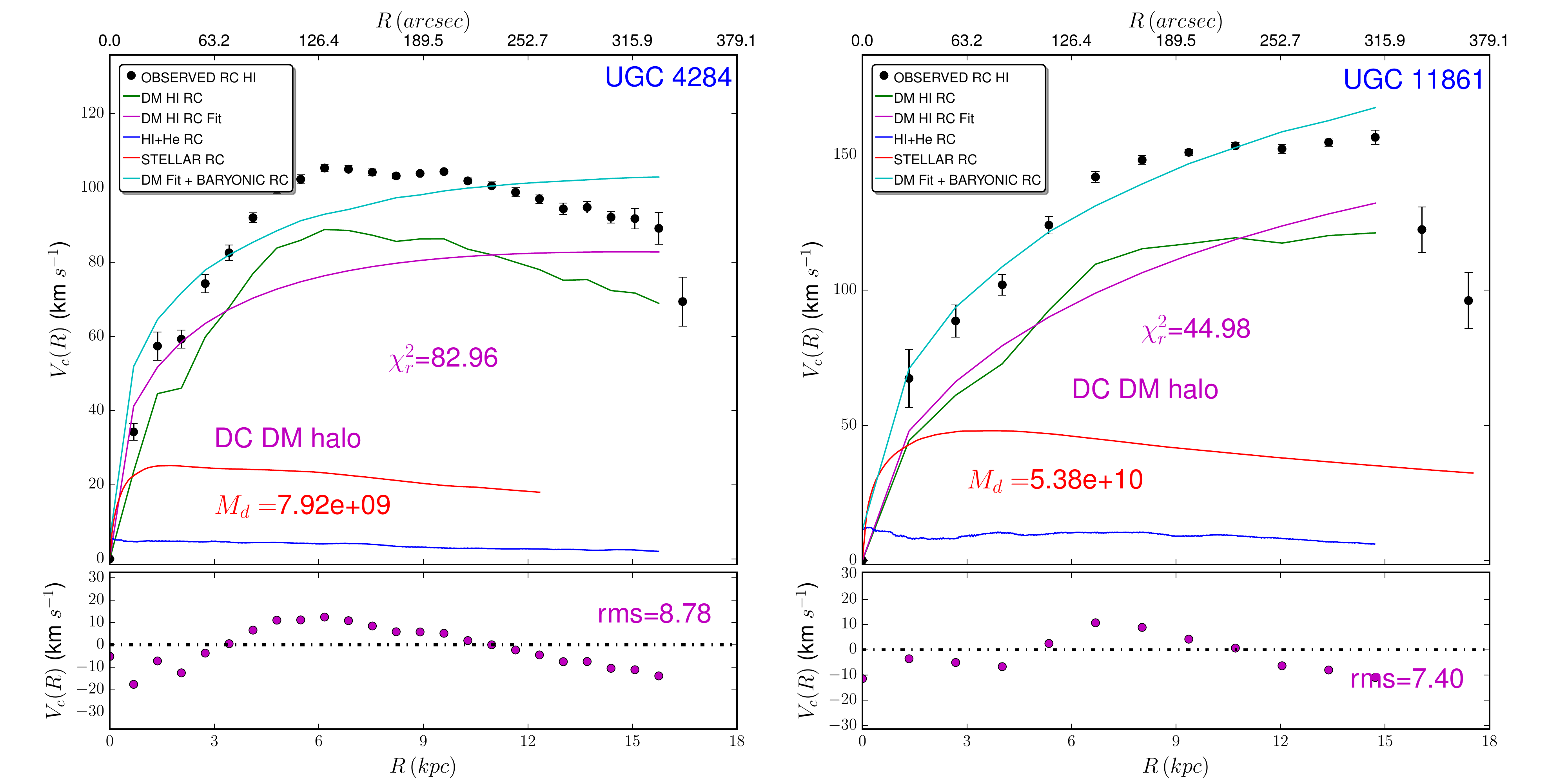}
\caption[f13.pdf]{{\it{Left}}: DCH fit of HI DM RC of UGC 4284. {\it{Right}}: DCH fit of HI DM RC of UGC 11861.
The number of degrees of freedom (DOF) are 22 for UGC 4284 and 10 for UGC 11861. The $\chi^2$ contours are displayed 
in figure~\ref{fig15}.
~\label{fig13}}
\end{figure*}

\begin{figure*}
\centering
\includegraphics[scale=0.4, angle=0]{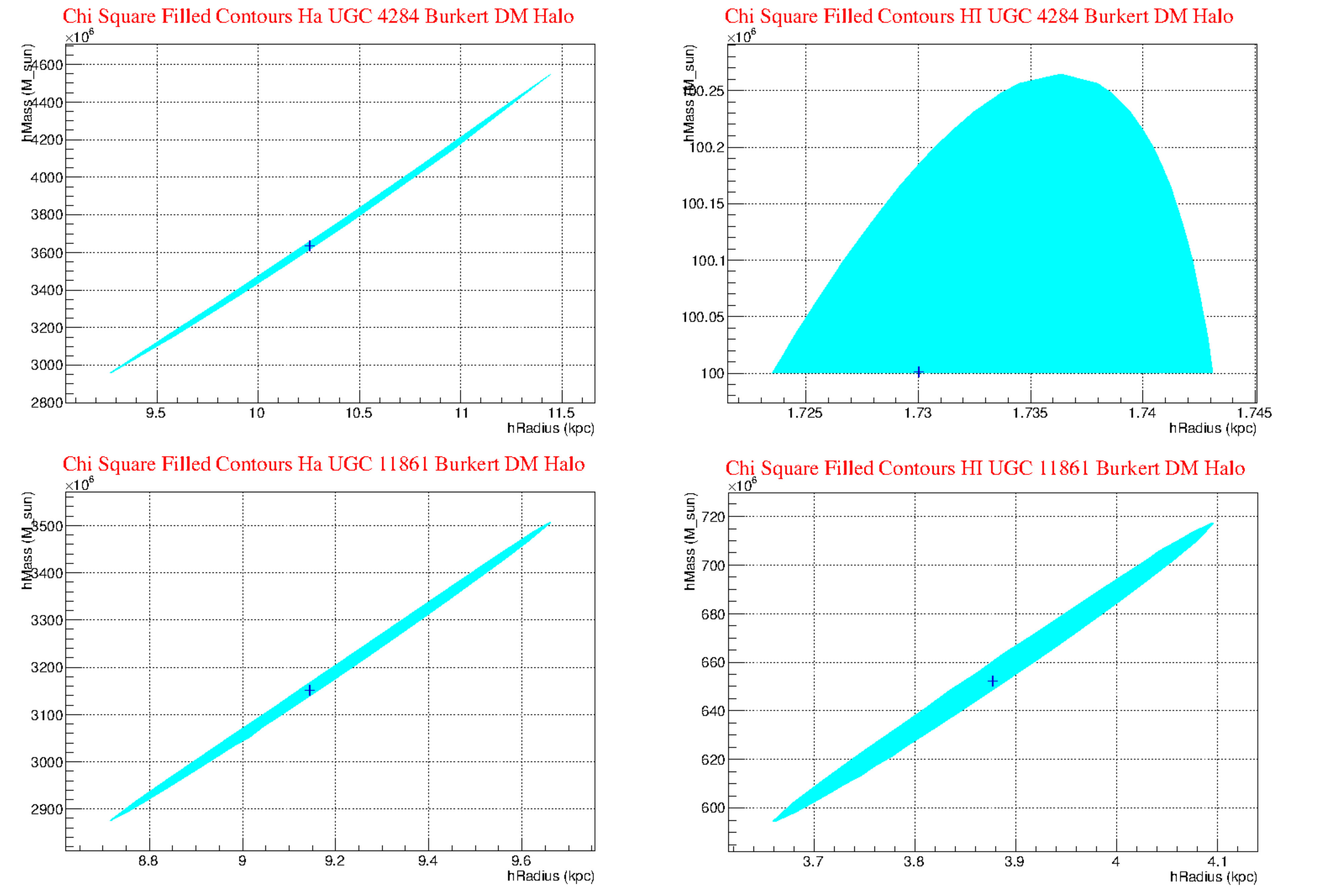}
\caption[f14.pdf]{$\chi^2$ filled contours at 70$\%$ confidence level. {\it First Panel}: H$\alpha$ DM RC of UGC 4284 BKH fit. 
{\it Second Panel}: HI DM RC of UGC 4284 BKH fit. {\it Third Panel}: H$\alpha$ DM RC of UGC 11861 BKH fit. {\it Fourth Panel}: 
HI DM RC of UGC 11861 BKH fit. The blue crosses represent the solutions found through the fitting procedure. 
~\label{fig14}}
\end{figure*}

\begin{figure*}
\centering
\includegraphics[scale=0.4, angle=0]{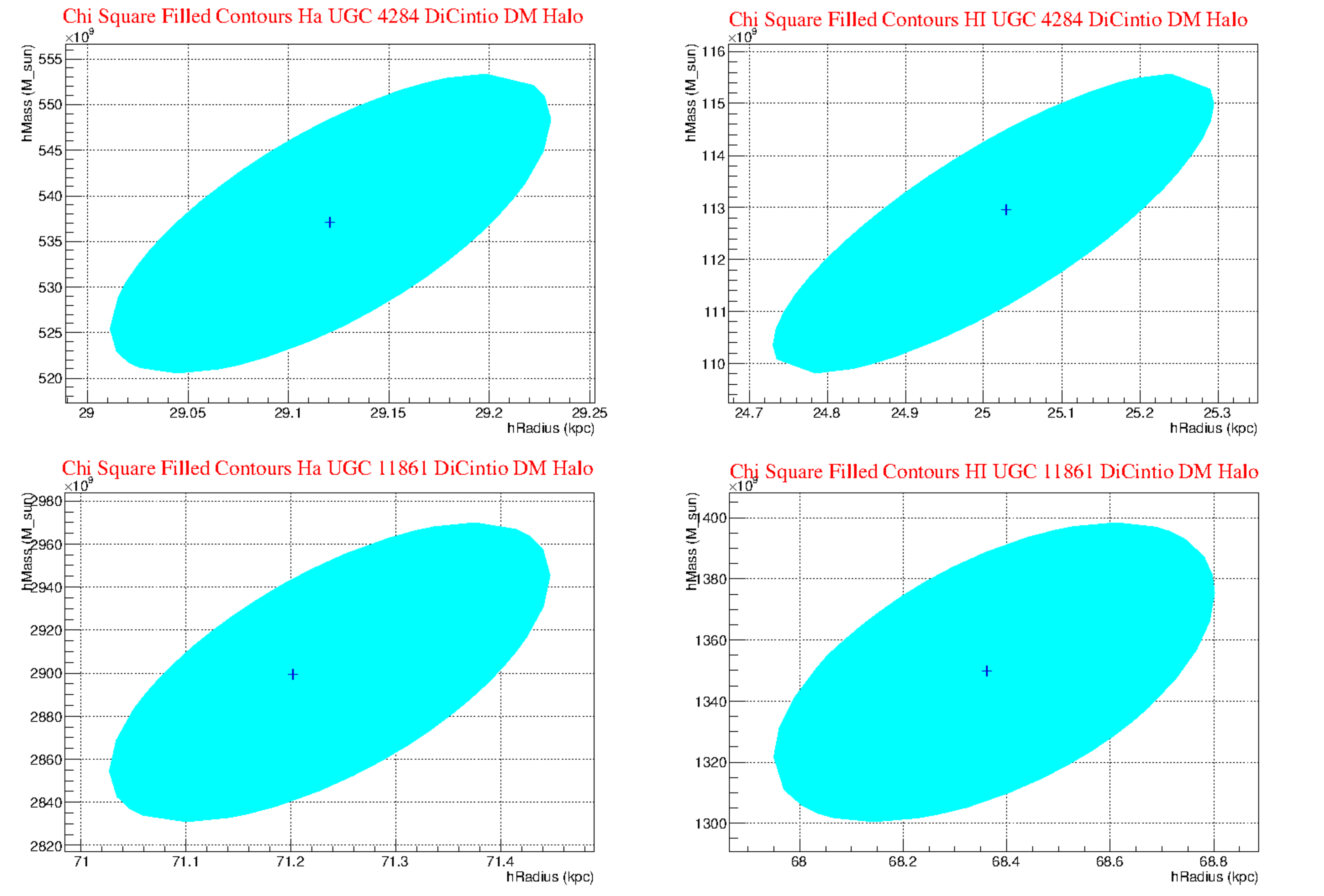}
\caption[f15.pdf]{$\chi^2$ filled contours at 70$\%$ confidence level. {\it First Panel}: H$\alpha$ DM RC of UGC 4284 DCH fit. 
{\it Second Panel}: HI DM RC of UGC 4284 DCH fit. {\it Third Panel}: H$\alpha$ DM RC of UGC 11861 DCH fit. {\it Fourth Panel}: 
HI DM RC of UGC 11861 DCH fit. The blue crosses represent the solutions found through the fitting procedure. 
~\label{fig15}}
\end{figure*}

\setcounter{table}{2}
\begin{table}
\centering 
\begin{minipage}{500mm}
\caption{H$\alpha$ DM RCs fit result for UGC 4284 (DOF=78).}
\begin{tabular}{@{}lccc@{}}
\hline
UGC 4284 NFW & $M_d+0\%$\footnote{$M_d=7.92 \times 10^9 M_{\odot}$} & $M_d-30\%$ & $M_d+30\%$.\\
\hline
$M_h (M_{\odot})$ & $7.07 \times 10^{12}$ & $3.75 \times 10^{12}$ & $3.98 \times 10^{12}$ \\
$R_h (kpc)$ & 97.90 & 76.33 & 65.26 \\ 
$\chi^2_r$\footnote{Reduced $\chi^2$} & 15.36 &  20.74 & 13.46 \\
Edm\footnote{Estimated distance from the minimum} & $1.45 \times 10^{9}$ & $1.30 \times 10^{9}$ & $1.70 \times 10^{9}$ \\
\hline
UGC 4284 BKH & $M_d+0\%$ & $M_d-30\%$ & $M_d+30\%$.\\
\hline
$M_h (M_{\odot})$ & $3.63 \times 10^{9}$ & $2.91 \times 10^{9}$ & $4.52 \times 10^{9}$ \\
$R_h$ (kpc) & 10.25 & 10.13 & 10.45\\ 
$\chi^2_r$ & 8.94 & 12.72 & 5.99\\
Edm & $1.22 \times 10^{-13}$ & $3.04 \times 10^{-13}$ & $1.74 \times 10^{-13}$\\
\hline
UGC 4284 DCH & $M_d+0\%$ & $M_d-30\%$ & $M_d+30\%$.\\
\hline
$M_h (M_{\odot})$ & $5.37 \times 10^{11}$ & $7.86 \times 10^{11}$ & $3.57 \times 10^{11}$ \\
$R_h (kpc)$ & 29.12 & 36.15 & 22.53\\ 
$\chi^2_r$ & 3.32 & 2.94 & 3.66 \\
Edm & $3.83 \times 10^{-15}$ & $7.09 \times 10^{-14}$ & $1.48 \times 10^{-18}$ \\
\hline
\hline
\end{tabular}
\end{minipage}
\end{table}

\setcounter{table}{3}
\begin{table}
\centering 
\begin{minipage}{500mm}
\caption{H$\alpha$ DM RCs fit result for UGC 11861 (DOF=43).}
\begin{tabular}{@{}lccc@{}}
\hline
\hline
UGC 11861 NFW & $M_d+0\%$\footnote{$M_d=5.38 \times 10^{10} M_{\odot}$} & $M_d-30\%$ & $M_d+30\%$.\\
\hline
$M_h (M_{\odot})$ & $2.93 \times 10^{12}$ & $7.05 \times 10^{12}$ & $4.79 \times 10^{12}$ \\
$R_h (kpc)$ & 54.26 & 97.91 & 65.27 \\ 
$\chi^2_r$\footnote{Reduced $\chi^2$} & 70.86 & 75.67 & 55.67 \\
Edm\footnote{Estimated distance from the minimum} & $9.62 \times 10^{8}$ & $1.37 \times 10^{9}$ & $5.58 \times 10^{10}$ \\
\hline
UGC 11861 BKH & $M_d+0\%$ & $M_d-30\%$ & $M_d+30\%$.\\
\hline
$M_h (M_{\odot})$ & $3.16 \times 10^{9}$ & $3.25 \times 10^{9}$ & $3.08 \times 10^{9}$ \\
$R_h$ (kpc) & 9.16 & 10.21 & 8.29\\ 
$\chi^2_r$ & 46.10 & 58.32 & 35.71\\
Edm & $1.40 \times 10^{-13}$ & $3.01 \times 10^{-12}$ & $9.10 \times 10^{-13}$\\
\hline
UGC 11861 DCH & $M_d+0\%$ & $M_d-30\%$ & $M_d+30\%$.\\
\hline
$M_h (M_{\odot})$ & $2.89 \times 10^{12}$ & $4.09 \times 10^{12}$ & $1.96 \times 10^{12}$  \\
$R_h (kpc)$ & 71.24 & 88.29 & 54.92\\ 
$\chi^2_r$ & 32.08 & 42.26 & 25.65\\
Edm & $2.40 \times 10^{-13}$ & $1.40 \times 10^{-14}$ & $6.03 \times 10^{-13}$ \\
\hline
\hline
\end{tabular}
\end{minipage}
\end{table}

\setcounter{table}{4}
\begin{table}
\centering 
\begin{minipage}{500mm}
\caption{HI DM RCs fit result for UGC 4284 (DOF=22).}
\begin{tabular}{@{}lccc@{}}
\hline
UGC 4284 NFW & $M_d+0\%$\footnote{$M_d=7.92 \times 10^9 M_{\odot}$} & $M_d-30\%$ & $M_d+30\%$.\\
\hline
$M_h (M_{\odot})$ & $4.16 \times 10^{12}$ & $3.64 \times 10^{12}$ & $2.94 \times 10^{12}$ \\
$R_h (kpc)$ & 97.90 & 97.90 & 79.03 \\ 
$\chi^2_r$\footnote{Reduced $\chi^2$} & 305.74 & 299.16 & 307.61 \\
Edm\footnote{Estimated distance from the minimum} & $4.77 \times 10^{8}$ & $4.52 \times 10^{8}$ & $2.61 \times 10^{9}$ \\
\hline
UGC 4284 BKH & $M_d+0\%$ & $M_d-30\%$ & $M_d+30\%$.\\
\hline
$M_h (M_{\odot})$ & $1.0 \times 10^{8}$ & $1.0 \times 10^{8}$ & $1.0 \times 10^{8}$ \\
$R_h$ (kpc) & 1.73 & 1.85 & 1.62\\ 
$\chi^2_r$ & 108.25 & 131.65 & 118.67\\
Edm & $1.34 \times 10^{-7}$ & $2.68 \times 10^{-7}$ & $1.67 \times 10^{-8}$\\
\hline
UGC 4284 DCH & $M_d+0\%$ & $M_d-30\%$ & $M_d+30\%$.\\
\hline
$M_h (M_{\odot})$ & $1.13 \times 10^{11}$ & $1.26 \times 10^{11}$ & $9.83 \times 10^{10}$ \\
$R_h (kpc)$ & 25.01 & 31.69 & 19.08\\ 
$\chi^2_r$ & 82.96 & 103.41 & 77.54\\
Edm & $3.18 \times 10^{-14}$ & $5.67 \times 10^{-14}$ & $2.67 \times 10^{-14}$ \\
\hline
\hline
\end{tabular}
\end{minipage}
\end{table}

\setcounter{table}{5}
\begin{table}
\centering 
\begin{minipage}{500mm}
\caption{HI DM RCs fit result for UGC 11861 (DOF=10).}
\begin{tabular}{@{}lccc@{}}
\hline
\hline
UGC 11861 NFW & $M_d+0\%$\footnote{$M_d=5.38 \times 10^{10} M_{\odot}$} & $M_d-30\%$ & $M_d+30\%$.\\
\hline
$M_h (M_{\odot})$ & $5.46 \times 10^{12}$ & $6.24 \times 10^{12}$ & $3.89 \times 10^{12}$ \\
$R_h (kpc)$ & 81.36 & 88.45 & 78.67 \\ 
$\chi^2_r$\footnote{Reduced $\chi^2$} & 97.35 & 95.34 & 92.67 \\
Edm\footnote{Estimated distance from the minimum} & $3.28 \times 10^{8}$ & $5.32 \times 10^{8}$ & $6.54 \times 10^{8}$ \\
\hline
UGC 11861 BKH & $M_d+0\%$ & $M_d-30\%$ & $M_d+30\%$.\\
\hline
$M_h (M_{\odot})$ & $6.53 \times 10^{8}$ & $6.12 \times 10^{8}$ & $6.94 \times 10^{8}$ \\
$R_h$ (kpc) & 3.87 & 4.01 & 3.75\\ 
$\chi^2_r$ & 37.60 & 38.35 & 36.93\\
Edm & $2.13 \times 10^{-13}$ & $2.53 \times 10^{-12}$ & $1.89 \times 10^{-12}$\\
\hline
UGC 11861 DCH & $M_d+0\%$ & $M_d-30\%$ & $M_d+30\%$.\\
\hline
$M_h (M_{\odot})$ & $1.35 \times 10^{12}$ & $1.70 \times 10^{12}$ & $1.01 \times 10^{12}$  \\
$R_h (kpc)$ & 68.38 & 85.26 & 52.22\\ 
$\chi^2_r$ & 44.98 & 54.95 & 48.08\\
Edm & $4.17 \times 10^{-13}$ & $3.59 \times 10^{-15}$ & $6.15 \times 10^{-13}$ \\
\hline
\hline
\end{tabular}
\end{minipage}
\end{table}

From the tables and figures presented in this section it is evident that the NFW DM halo does not fit either the H$\alpha$ or the HI DM RCs of UGC 4284 and UGC 11861.
The Edm indicates that the solution is far from the expected global minimum and this fact is also confirmed by the non existence of the $\chi^2$ filled contours.
The BK DM halo fits the H$\alpha$ and HI DM RCs of UGC 4284 and UGC 11861 better than NFW, even if the $\chi^2$ filled contours are narrow and for the HI DM RC
of UGC 4284 the final halo mass value coincide with the lower halo mass limit allowed by our fitting procedure. 

The DC DM halo seems to perform better than BK DM halo as indicated by the $\chi^2$ values, the $\chi^2$ filled contours and the Edm values, even if in the case of the
DM RC of UGC 11861 the $\chi^2$ values for the BK DM halo are lower than the DC DM halo. The DM masses originated from the fitting of the DC DM halo are in general a factor
$10^2-10^3$ times greater than the DM halo masses generated by the fitting of the BK DM halo. In the case of the HI DM RC of UGC 11861 the DM masses of the DC DM halo are a 
factor of $10^4$ greater than BK DM halo, even if in this case the BK DM halo accomplish better than the DC DM halo. Considering the range of disk stellar masses analysed in
this study (see section~\ref{hia_rcs}) and DM halo masses obtained throughout this analysis for the DC DM halo our results reproduced closely the results presented by 
\citet{DiCintio2014b} (see figure 4 of the quoted article), in fact the inner slope outcome from the DM RC fitted in this work with the DC DM halo are inside the interval 
$\left[-0.6\,,\,-1.2\right]$, therefore the results of the fits performed with the DC DM halos give in all cases a cuspy DM profile. This fact could in part justify the higher
DM halo masses, with respect to BK DM halo, needed to reproduce the DM RCs considered in this work using the DC DM halo. In general the fitting errors are about $0.01\%$ for 
the DM halos radii and $0.1\%$ for the DM halos masses, in the case where the DM halos fits succeeded.

In Appendix A we additionally test the Einasto \citep{Einasto1965} DM halo (HE) and the Stadel \citep{Stadel2009} DM halo (HSD) as feasible options to model the H$\alpha$ and HI 
DM RCs of UGC 4284 and UGC 11861. Our purpose is to compare the outcome from these two DM halos with the other results encountered in this research to establish if the analysed 
RCs are best reproduced by cuspy or core profiles. As mentioned in Appendix A we regarded the HE and HSD DM halos as two free parameters DM halos fixing the third parameters for 
both halos in order to have the same degrees of freedom of the other DM halos considered in this work. The results showed in the tables~A1 and A2 clearly determine that the cored 
HE and HSD DM halos fits better than BK DM halo and also than DC DM halo practically in all cases, with the exception of the DM H$\alpha$ RC of UGC 4284 where the DC DM halo 
produces better results. In summary our main result is that the cored DM profiles HE and HSD better represent the DM content of UGC 4284 and UGC 11861.}


\section{Discussion}

{ The $\Lambda$CDM  ($\Lambda$ Cold Dark Matter) paradigm, predicts cuspy density DM profiles in the centre of galaxies, whereas observational tests, using RCs 
analysis of spirals, irregular, dwarf, and low surface brightness galaxies found that cored density DM profiles better represent the central part of observed galaxies. 
The disagreement between the $\Lambda$CDM scenario and the observational results, led to consider possible modifications to the $\Lambda$CDM scheme.
One possible refinement of the $\Lambda$CDM paradigm consists on the incorporation of baryonic physics on small scales ($<$ 1 kpc). The basic idea to transform a cuspy central 
density profile to a cored one, is to remove baryonic matter from the central kpc of galaxies through some perturbation to their gravitational potential. Supernova explosions and 
driven-gas outflows, for instance, could be very efficient processes to obtain a central DM core in the centre of galaxies. Other possible mechanisms are gas and metals cooling 
and star formation. Some numerical and analytical attempts have been successful to achieve the desired DM cored central profile \citep{Governato2010, Governato2012, deSouza2011, 
DiCintio2014a}.

Regardless of all the considerable observational and theoretical endeavours to contribute to the solution of the cuspy/core controversy, the debate is still open and in this work 
we attempt to discriminate between cuspy and core DM halos through DM RCs fitting after subtracting from the observed RCs the stellar and gas rotational components.
We consider high resolution H$\alpha$ and HI RCs of two irregular galaxies, UGC 4284 and UGC 11861, available in the literature, to 
examine the mass content of both objects by subtracting to the observed H$\alpha$ the stellar rotation velocity and to the HI RCs both the stellar and gas rotation
velocity. We obtained the stellar rotational component through stellar population synthesis of the disk of these two galaxies.} We recognise that in the case of both galaxies, 
once subtracted the determined stellar and gas rotation velocity, it is missing more than 50$\%$ of mass, 
for a stellar disk mass of $M_d=7.92 \times 10^9 M_{\odot}$ in the case of UGC 4284 and more than 40$\%$ of mass for a stellar disk mass of $M_d=5.38 \times 
10^{10} M_{\odot}$ in the case of UGC 11861. { We explored the possibility that DM could account for the missing mass in these galaxies, and we fitted the derived DM RCs with 
five DM models, namely the NFW DM halo, the BK DM halo, the DC DM halo, the HE DM halo and the HSD DM halo, to estimate the assumed DM contribution to the total mass of UGC 4284 
and UGC 11861.}

{ The cuspy DC DM halo fit better the H$\alpha$ DM RC of UGC 4284 than any other DM halos used in this study, in all the other cases, say HI RC of UGC 4284 and 
H$\alpha$ and HI RCs of UGC 11861, the cored DM models give a better fit to the DM RCs of both galaxies than the cuspy DM profiles. In more detail the HE and HSD DM 
halos with two free parameters perform better than the BK DM halos except for the H$\alpha$ DM RC of UGC 4284 with stellar disk mass $M_d+30\%$. The HE DM halo achieve 
better results than the HSD DM halo for the HI DM RCs of both galaxies and also for the DM H$\alpha$ RC of UGC 11861 with stellar disk mass $M_d+0\%$ and $M_d+30\%$. 
The HSD DM halo give better results than the HE for the DM H$\alpha$ RC of UGC 4284 and DM RC of UGC 11861 with stellar disk mass $M_d-30\%$. As mentioned in section~\ref{io_dm} 
the cuspy result we obtain in the case of the H$\alpha$ DM RC of UGC 4284 is in accordance with the numerical study of \citet{DiCintio2014b} (see figure 4 of that article), 
given that our range of stellar, DM halo masses and inner slopes coincide with those presented in the quoted figure. Another important outcome of this study is the fact that 
the $M/L$ ratio is not constant in none of the two studied galaxies (see figures~\ref{fig3} and \ref{fig4}).}

At this point we can answer, at least for UGC 4284 and UGC 11861, to the question we asked in the Introduction, i.e., if the DM profile slope could change 
substantially obtaining the stellar disk from stellar population studies instead of assuming a constant M/L as previous studies done in the past (see 
Introduction and references therein). From the results outlined above it seems that in the case of UGC 4284 and UGC 11861, the answer to this question is a 
negative one. In the future we plan to repeat this study for other high resolution RCs, in order to asses this question for a larger number of galaxies.

\section{Conclusions}

{ In this study we analyse the inner and outer mass distribution of the two irregular galaxies UGC 4284 and UGC 11861, by means of high resolution H$\alpha$ and HI 
RCs available in the literature. Additionally, we constrained the stellar content of UGC 4284 and UGC 11861 throughout stellar population synthesis studies, 
applied to the photometric observations of these two galaxies in several filters, we also derive $HI$+$He$ RCs for the HI observed RCs of both galaxies and we subtracted 
the derived stellar disk from the observed H$\alpha$ RCs and the stellar and gas disk from the observed HI RCs. 

We fitted the determined DM RCs for both galaxies, examining the NFW, BK, DC, HE and HSD DM models. The principal results of this research can be summarized according to 
the following points:\\

\begin{enumerate}
 \item Cored DM models better fit the observed shape of the two DM H$\alpha$ and HI RCs considered in this study, with respect to cuspy DM models,
 except for the H$\alpha$ DM RC of UGC 4284 where the cuspy DC DM halo is the preferable model.\\
 
 \item Cored exponential DM model Einasto reproduced better the DM HI RCs of UGC 4284, and UGC 11861, than the other models tested in this paper.\\
 
 \item Cored exponential DM model Einasto reproduced better than all the other models the DM H$\alpha$ RCs of UGC 11861 for stellar masses $M_d+0\%$ and $M_d+30\%$.\\
 
 \item The cored Stadel DM model, in the case of the H$\alpha$ DM RC of UGC 4284, performed better than Einasto DM models for all the stellar masses considered.\\
 
 \item The cored Stadel DM model, in the case of the H$\alpha$ DM RC of UGC 11861 and a stellar mass $M_d-30\%$, achieved better results than Einasto DM models.\\
 
\end{enumerate}}

\section*{Acknowledgments}

RP acknowledges CNPq's financial support for postdoctoral fellowship at LNA.
EMG acknowledges CONACYT's financial support for postdoctoral fellowship at INAOE.
We also acknowledge the DGAPA-UNAM project IN 108912 and CONACyT-CB-2007/82389Â project.
We acknowledge Janine Van Eymeren for providing us the HI total density map of both galaxies. 

\bibliographystyle{mn2e}
\bibliography{biblio}

\appendix

\section[appx1]{Einasto and Stadel DM RC fit results}

{ We also consider the Einasto (HE) \citep{Einasto1965} and the Stadel \citep{Stadel2009} (HSD) DM velocity profiles in our endeavour to reproduce the H$\alpha$
and HI DM RCs of UGC 4284 and UGC 11861. The motivation to employ the HE and HSD DM halos are emphasized below, even if at the moment it does not exists a physical
justification of these two DM profiles. During the fitting process we fixed the third parameter of both DM halos, therefore the number of degrees of freedom
is the same of the other DM halos examined in this work.

The HE was introduced for the first time by \citet{Einasto1965} to build a model of the distribution of stars in the Galaxy.
\citet{Navarro2004, Navarro2010} established that the HE fits the DM density profiles produced in the framework of $\Lambda$CDM cosmological numerical 
simulations better than the NFW profile. \citet{Chemin2011} used for the first time the HE DM models to fit 34 nearby galaxies, determining that the HE DM halo
represented better the observed RCs of the studied galaxies than the Pseudo-Isothermal DM model.

The HE represents a three parameters model in the form of an exponential-like profile and also incorporates cusped 
and cored DM halos. We used the approximation of \citet{Ciotti1999} for the dimensionless scale factor $d_n$.

\setcounter{table}{0}
\begin{table*}
\centering 
\begin{minipage}{500mm}
\caption{H$\alpha$ DM RC fit results (UGC 4284 DOF=78 AND UGC 11861 DOF=43).}
\begin{tabular}{@{}lccc@{}}
\hline
\hline
UGC 4284 EINASTO & $M_d+0\%$ & $M_d-30\%$ & $M_d+30\%$.\\
\hline
$M_h (M_{\odot})$ & $1.22 \times 10^{12}$ & $1.26 \times 10^{12}$ & $1.22 \times 10^{12}$\\
$R_h (kpc)$ & 100 & 100 & 100\\
$n_h$ & 3.16 & 2.92 & 3.37 \\
$\chi^2_r$ & 4.43 & 4.04 & 4.96 \\
Edm & $8.37 \times 10^{-8}$ & $2.40 \times 10^{-7}$ & $4.10 \times 10^{-7}$ \\
\hline
UGC 4284 STADEL & $M_d+0\%$ & $M_d-30\%$ & $M_d+30\%$.\\
\hline
$M_h (M_{\odot})$ & $8.70 \times 10^{11}$ & $1.14 \times 10^{12}$ & $5.17 \times 10^{11}$ \\
$R_h (kpc)$ & 47.13 & 59.65 & 33.73\\ 
$\lambda_h$ & 3.79 & 3.80 & 3.81\\
$\chi^2_r$ & 6.13 & 7.10 & 8.77\\
Edm & $1.07 \times 10^{-11}$ & $1.61 \times 10^{-14}$ & $6.39 \times 10^{-9}$ \\
\hline
UGC 11861 EINASTO & $M_d+0\%$ & $M_d-30\%$ & $M_d+30\%$.\\
\hline
$M_h (M_{\odot})$ & $1.19 \times 10^{11}$ & $8.27 \times 10^{10}$ & $1.81 \times 10^{11}$\\
$R_h (kpc)$ & 14.92 & 12.85 & 18.12\\
$n_h$ & 1.13 & 0.92 & 1.39 \\
$\chi^2_r$ & 6.18 & 10.04 & 3.74 \\
Edm & $3.81 \times 10^{-14}$ & $1.68 \times 10^{-13}$ & $3.88 \times 10^{-16}$ \\
\hline
UGC 11861 STADEL & $M_d+0\%$ & $M_d-30\%$ & $M_d+30\%$.\\
\hline
$M_h (M_{\odot})$ & $5.27 \times 10^{11}$ & $4.61 \times 10^{11}$ & $9.86 \times 10^{11}$ \\
$R_h (kpc)$ & 31.23 & 31.22 & 41.89\\ 
$\lambda_h$ & 3.65 & 3.59 & 3.75\\
$\chi^2_r$ & 8.05 & 8.02 & 8.96\\
Edm & $1.74 \times 10^{-13}$ & $2.70 \times 10^{-16}$ & $4.11 \times 10^{-8}$ \\
\hline
\hline
\end{tabular}
\end{minipage}
\end{table*}

\setcounter{table}{1}
\begin{table*}
\centering 
\begin{minipage}{500mm}
\caption{HI DM RC fit results (UGC 4284 DOF=22 AND UGC 11861 DOF=10).}
\begin{tabular}{@{}lccc@{}}
\hline
\hline
UGC 4284 EINASTO & $M_d+0\%$ & $M_d-30\%$ & $M_d+30\%$.\\
\hline
$M_h (M_{\odot})$ & $1.82 \times 10^{10}$ & $1.59 \times 10^{10}$ & $2.06 \times 10^{10}$\\
$R_h (kpc)$ & 5.42 & 5.23 & 5.64\\
$n_h$ & 0.80 & 0.73 & 0.87 \\
$\chi^2_r$ & 5.90 & 5.74 & 6.10 \\
Edm & $1.68 \times 10^{-15}$ & $1.59 \times 10^{-15}$ & $5.99 \times 10^{-16}$ \\
\hline
UGC 4284 STADEL & $M_d+0\%$ & $M_d-30\%$ & $M_d+30\%$.\\
\hline
$M_h (M_{\odot})$ & $7.41 \times 10^{10}$ & $6.44 \times 10^{10}$ & $8.02 \times 10^{10}$ \\
$R_h (kpc)$ & 12.18 & 11.60 & 11.84\\ 
$\lambda_h$ & 3.82 & 3.77 & 3.73\\
$\chi^2_r$ & 25.30 & 24.33 & 17.95\\
Edm & $3.52 \times 10^{-12}$ & $4.78 \times 10^{-12}$ & $1.68 \times 10^{-14}$ \\
\hline
UGC 11861 EINASTO & $M_d+0\%$ & $M_d-30\%$ & $M_d+30\%$.\\
\hline
$M_h (M_{\odot})$ & $5.70 \times 10^{10}$ & $5.16 \times 10^{10}$ & $6.54 \times 10^{10}$\\
$R_h (kpc)$ & 9.20 & 10.30 & 11.40\\
$n_h$ & 1.12 & 1.10 & 1.15 \\
$\chi^2_r$ & 19.50 & 17.25 & 20.51 \\
Edm & $1.26 \times 10^{-8}$ & $1.30 \times 10^{-9}$ & $1.17 \times 10^{-9}$ \\
\hline
UGC 11861 STADEL & $M_d+0\%$ & $M_d-30\%$ & $M_d+30\%$.\\
\hline
$M_h (M_{\odot})$ & $3.65 \times 10^{11}$ & $3.31 \times 10^{11}$ & $4.0 \times 10^{11}$ \\
$R_h (kpc)$ & 26.60 & 27.04 & 26.18\\ 
$\lambda_h$ & 3.79 & 3.79 & 3.79\\
$\chi^2_r$ & 26.70 & 31.38 & 27.98\\
Edm & $2.06 \times 10^{-14}$ & $1.44 \times 10^{-13}$ & $1.21 \times 10^{-14}$ \\
\hline
\hline
\end{tabular}
\end{minipage}
\end{table*}

The HSD have been proposed by \cite{Stadel2009} accomplishing several $\Lambda$CDM Galactic mass DM halo numerical 
simulations, the authors concluded that the HSD better reproduced the shape of the simulated DM halos density profiles also if compared with the HE. The 
HSD is an exponential-like profile with three free parameters, but according to \citet{Stadel2009} can also be regarded as a DM density profile with two 
free parameters with functional form:

\[
 \rho(r)=\rho_0 e^{-\lambda\left[\ln{(1+r/R_{\lambda})}\right]^2},
\]

where $R_{\lambda}$ is the HSD scale radius and $\lambda$ is a parameter that governs the shape of the HSD. In this study we also provide an exact derivation 
of the HSD velocity profile in Appendix B. In particular the relation (B10) of Appendix B, is a tabulated integral from \citet{gradshteyn2007}. 
The relation (B19) of the same Appendix B, is the expression we used in this investigation to fit the DM RCs of UGC 4284 and UGC 11861 employing the HSD.
The results of the HE and HSD fits are listed in table~A1 and table~A2.}

\section[appx2]{Circular velocity derivation for the Stadel DM model}

The spatial density relation for the HSD halo density profile is expressed by the relation:

\begin{equation}
\rho(r)=\rho_0 e^{-\lambda\left[\ln\left(1+\frac{r}{R_{\lambda}}\right)\right]^2}.  
\end{equation}

\noindent We can write this spatial density in the following form:

\begin{equation}
\rho(r)=\rho_0\left[e^{\ln\left(1+\frac{r}{R_{\lambda}}\right)^{-\lambda}}\right]^{\ln\left(1+\frac{r}{R_{\lambda}}\right)},  
\end{equation}

\noindent using exponential properties we obtain the relation:

\begin{equation}
\rho(r)=\rho_0\left[\frac{1}{\left(1+\frac{r}{R_{\lambda}}\right)^{\lambda}}\right]^{\ln\left(1+\frac{r}{R_{\lambda}}\right)} 
\end{equation}

\noindent thus we can use this form of the spatial density to obtain the mass density up to a radius $r$, with the relation:

\begin{equation}
M(r)=4\pi\int_0^r \rho(s) s^{2} ds 
\end{equation}

\noindent substituting in this relation we obtain:

\begin{equation}
M(r)=4\pi\rho_0\int_0^r\left[\frac{1}{\left(1+\frac{s}{R_{\lambda}}
\right)^{\lambda}}\right]^{\ln\left(1+\frac{s}{R_{\lambda}}\right)}s^{2} ds   
\end{equation}

\noindent with the substitution $e^u=\frac{1}{\left(1+\frac{s}{R_{\lambda}}\right)^{\lambda}} \Rightarrow 
e^{\frac{u}{\lambda}}=\frac{1}{\left(1+\frac{s}{R_{\lambda}}\right)}$ we obtain $s=R_{\lambda}\left(\frac{1-e^{\frac{u}{\lambda}}}
{e^{\frac{u}{\lambda}}}\right) \Rightarrow s=R_{\lambda}\left(e^{-\frac{u}{\lambda}}-1\right)$ and $u=-\lambda \ln\left(1+\frac{s}
{R_{\lambda}}\right) \Rightarrow du=-\lambda\frac{d}{ds}\ln\left(1+\frac{s}{R_{\lambda}}\right) \Rightarrow du=-\lambda \, 
\frac{\frac{1}{R_{\lambda}}}{\left(1+\frac{s}{R_{\lambda}}\right)} \, ds \Rightarrow du=-\frac{\lambda}{R_{\lambda}} 
e^{\frac{u}{\lambda}} \, ds \Rightarrow ds=-\frac{R_{\lambda}}{\lambda} e^{-\frac{u}{\lambda}} \, du$.

\noindent For the limits of integration we have that $u=\ln\left[\frac{1}{\left(1+\frac{s}{R_{\lambda}}\right)^{\lambda}}\right]$ we obtain $u=0$ 
for $r=0$ and $u=\ln\left[\frac{1} {\left(1+\frac{r}{R_{\lambda}}\right)^{\lambda}}\right]$ for $r=r$. 

\noindent It is important to note that the upper integration limit is negative given that $u=\ln\left[\left(1+\frac{r}{R_{\lambda}}\right)^{-\lambda}\right] 
\Rightarrow u=-\lambda\ln\left[\left(1+\frac{r}{R_{\lambda}}\right)\right]$ with $(\lambda \,, r \,, R_{\lambda}) > 0$.

\noindent We proceed in the calculation of the integral substituting all the derived quantities in the above relation obtaining:

\begin{equation}
M(r)=-\frac{4\pi \rho_0 R^3_{\lambda}}{\lambda} \int^{(u < 0)}_0 e^{-\frac{s^2}{\lambda}}
\left(e^{-\frac{s}{\lambda}}-1\right)^2 e^{-\frac{s}{\lambda}}  ds  
\end{equation}

\noindent After simplifying this relation and making the square we obtain the following relation:

\begin{equation}
M(r)=\frac{4\pi \rho_0 R^3_{\lambda}}{\lambda} \int^0_{(u < 0)} e^{-\frac{s^2}{\lambda}} 
\left(e^{-\frac{3s}{\lambda}}+e^{-\frac{s}{\lambda}}-2e^{-\frac{2s}{\lambda}}\right)  ds  
\end{equation}

\noindent multiplying and simplifying we get the expression:

\begin{equation}
M(r)=\frac{4\pi \rho_0 R^3_{\lambda}}{\lambda} \int^0_{(u < 0)} \left[e^{-\left(\frac{s^2+3s}{\lambda}\right)}+
e^{-\left(\frac{s^2+s}{\lambda}\right)}-2e^{-\left(\frac{s^2+2s}{\lambda}\right)}\right] ds  
\end{equation}

\noindent naming the first integral in the above expression as $A(s)=\int^0_{(u < 0)} e^{-\left(\frac{s^2+3s}{\lambda}\right)} ds$, the 
second $B(s)=\int^0_{(u < 0)} e^{-\left(\frac{s^2+s}{\lambda}\right)} ds$ and the third integral $C(s)=\int^0_{(u < 0)} e^{-\left(
\frac{s^2+2s}{\lambda}\right)} ds$ we can rewrite our mass integral in the following manner:

\begin{equation}
M(r)=\frac{4\pi \rho_0 R^3_{\lambda}}{\lambda} \left[A(s)+B(s)-2C(s)\right]  
\end{equation}

\noindent At this point considering each integral separately and considering the indefinite integral given by the expression:

\begin{equation}
\int e^{-\left(ax^2+2bx+c\right)} dx=\frac{1}{2}\sqrt{\frac{\pi}{a}}\,\exp{\left(\frac{ac-b^2}{a}\right)}
\erf{\left[\sqrt{a}x+\frac{b}{\sqrt{a}}\right]}  
\end{equation}

where the $erf(x)$ is given by relation $erf(x)=\frac{2}{\sqrt{\pi}}\int_0^x e^{-t^2} dt$,

in the case of the first integral $A(s)$, we obtain $a=\frac{1}{\lambda}$; $b=\frac{3}{2\lambda}$.
For the second integral $B(s)$ in a similar way we get $a=\frac{1}{\lambda}$; $b=\frac{1}{2\lambda}$ and for the 
third integral we obtain $a=\frac{1}{\lambda}$; $b=\frac{1}{\lambda}$. All the integrals have $c=0$.

Thus the previously indefinite integrals $A(s)$, $B(s)$ and $C(s)$ transform according to the relation:

\begin{equation}
A(s)=\int e^{-\left(\frac{s^2+3s}{\lambda}\right)} ds=\frac{\sqrt{\pi\lambda}}{2}\,\exp{\left(-\frac{9}{4\lambda}\right)}
\erf{\left[\sqrt{\frac{1}{\lambda}}s+\frac{3}{2\sqrt{\lambda}}\right]}    
\end{equation}

\begin{equation}
B(s)=\int e^{-\left(\frac{s^2+s}{\lambda}\right)} ds=\frac{\sqrt{\pi\lambda}}{2}\,\exp{\left(-\frac{1}{4\lambda}\right)}
\erf{\left[\sqrt{\frac{1}{\lambda}}s+\frac{1}{2\sqrt{\lambda}}\right]}    
\end{equation}

\begin{equation}
C(s)=\int e^{-\left(\frac{s^2+2s}{\lambda}\right)} ds=\frac{\sqrt{\pi\lambda}}{2}\,\exp{\left(-\frac{1}{\lambda}\right)}
\erf{\left[\sqrt{\frac{1}{\lambda}}s+\frac{1}{\sqrt{\lambda}}\right]}   
\end{equation}

now, we calculate the definite integrals setting $u=-v$ where $v > 0$ with the reminder that $erf(-x)=-erf(x)$ we obtain the following results:

\begin{equation}\nonumber
\begin{split}
\int^0_{(-v < 0)} e^{-\left(\frac{s^2+3s}{\lambda}\right)} ds=A(0)-A(-v)=\frac{\sqrt{\pi\lambda}}{2}\,\exp{\left(-\frac{9}{4\lambda}\right)} \\
\times\left\{\erf{\left[\sqrt{\frac{1}{\lambda}}v-\frac{3}{2\sqrt{\lambda}}\right]}+\erf{\left[\frac{3}{2\sqrt{\lambda}}\right]}\right\}  
\end{split}
\end{equation}

\begin{equation}\nonumber
\begin{split}
\int^0_{(-v < 0)} e^{-\left(\frac{s^2+s}{\lambda}\right)} ds=B(0)-B(-v)=\frac{\sqrt{\pi\lambda}}{2}\,\exp{\left(-\frac{1}{4\lambda}\right)} \\
\times\left\{\erf{\left[\sqrt{\frac{1}{\lambda}}v-\frac{1}{2\sqrt{\lambda}}\right]}+\erf{\left[\frac{1}{2\sqrt{\lambda}}\right]}\right\}  
\end{split}
\end{equation}

\begin{equation}\nonumber
\begin{split}
\int^0_{(-v < 0)} e^{-\left(\frac{s^2+2s}{\lambda}\right)} ds=C(0)-C(-v)=\frac{\sqrt{\pi\lambda}}{2}\,\exp{\left(-\frac{1}{\lambda}\right)} \\
\times\left\{\erf{\left[\sqrt{\frac{1}{\lambda}}v-\frac{1}{\sqrt{\lambda}}\right]}+\erf{\left[\frac{1}{\sqrt{\lambda}}\right]}\right\}  
\end{split}
\end{equation}

Let us define the part involving the erf functions in the above expressions in the following way:

\begin{equation}
I_{erf1}(v, \lambda)=\left\{\erf{\left[\sqrt{\frac{1}{\lambda}}v-\frac{3}{2\sqrt{\lambda}}\right]}
+\erf{\left[\frac{3}{2\sqrt{\lambda}}\right]}\right\}  
\end{equation}

\begin{equation}
I_{erf2}(v, \lambda)=\left\{\erf{\left[\sqrt{\frac{1}{\lambda}}v-\frac{1}{2\sqrt{\lambda}}\right]}
+\erf{\left[\frac{1}{2\sqrt{\lambda}}\right]}\right\}  
\end{equation}

\begin{equation}
I_{erf3}(v, \lambda)=\left\{\erf{\left[\sqrt{\frac{1}{\lambda}}v-\frac{1}{\sqrt{\lambda}}\right]}
+\erf{\left[\frac{1}{\sqrt{\lambda}}\right]}\right\}  
\end{equation}

at this point we can rewrite the HSD mass function in the following form:

\begin{equation}
M(r)=\frac{2\pi^{3/2} \rho_0 R^3_{\lambda}}{\lambda^{1/2}}\left[e^{-\frac{9}{4\lambda}}I_{erf1}(v, \lambda)
+e^{-\frac{1}{4\lambda}}I_{erf2}(v, \lambda)-2e^{-\frac{1}{\lambda}}I_{erf3}(v, \lambda)\right]  
\end{equation}

where $-v=\ln\frac{1}{\left(1+\frac{r}{R_{\lambda}}\right)^{\lambda}}$ and the circular velocity is given by $V_c(r)=GM(r)/r$ so the above relation becomes:

\begin{equation}
V^2_c(r)=\frac{2\pi^{3/2} G \rho_0 R^3_{\lambda}}{\lambda^{1/2}{r}}\left[e^{-\frac{9}{4\lambda}}I_{erf1}(v, \lambda)
+e^{-\frac{1}{4\lambda}}I_{erf2}(v, \lambda)-2e^{-\frac{1}{\lambda}}I_{erf3}(v, \lambda)\right]   
\end{equation}

we can express the cicular velocity expression as a function of the initial mass of the system remembering that $M_{\lambda}=\rho_0 V_{\lambda}$, 
where $V_{\lambda}=\frac{4}{3}\pi R^3_{\lambda}$ is the volume of a sphere of radius $r$ containing the mass $M$ for definition, so we obtain:

\begin{equation}
V^2_c(r)=\frac{3 M_{\lambda}\pi^{1/2} G}{2\lambda^{1/2}{r}}\left[e^{-\frac{9}{4\lambda}}I_{erf1}(v, \lambda)
+e^{-\frac{1}{4\lambda}}I_{erf2}(v, \lambda)-2e^{-\frac{1}{\lambda}}I_{erf3}(v, \lambda)\right]  
\end{equation}

where $M_{\lambda}=\frac{4}{3}\pi R^3_{\lambda} \rho_0$ is the mass enclosing the initial density $\rho_0$ inside a radius $R_{\lambda}$ and $G$ is
the gravitational constant expressed in astrophysical units ($M^{-1}_{\odot}$ kpc $km^2 s^{-2}$).

\end{document}